\documentclass[10pt,aps,pre,twocolumn,superscriptaddress]{revtex4-1}

\usepackage{amsmath,amsthm}
\usepackage{amssymb}
\usepackage{amscd}
\usepackage{wasysym}
\usepackage[ansinew]{inputenc}
\usepackage[T1]{fontenc}
\usepackage{ae,aecompl}
\usepackage{sidecap}
\usepackage{hyperref}

    \usepackage[pdftex]{graphicx}
    \DeclareGraphicsExtensions{.pdf}

\usepackage{color}
\definecolor{red}{rgb}{1,0,0}
\definecolor{lred}{rgb}{1,0.7,0.7}
\definecolor{lgreen}{rgb}{0.7,1,0.7}

\usepackage{colortbl}

\newcommand{\be}{\begin{equation}}
\newcommand{\ee}{\end{equation}}
\newcommand{\bea}{\begin{eqnarray}}
\newcommand{\eea}{\end{eqnarray}}
\newcommand{\nn}{\nonumber}

\renewcommand{\vec}[1]{\boldsymbol{#1}}

\newcommand{\OO}{\mathcal{O}}

\newtheorem{definition}{Definition}

\begin{document}
\bibliographystyle{apsrev}

\title{Network susceptibilities: theory and applications}

\author{Debsankha Manik}
\thanks{These authors contributed equally to this work.}
\affiliation{Network Dynamics, Max Planck Institute for Dynamics and Self-Organization (MPIDS), 37077 G\"ottingen, Germany}

\author{Martin Rohden}
\thanks{These authors contributed equally to this work.}
\affiliation{Department of Physics and Earth Sciences, Jacobs University, 28759 Bremen, Germany}

\author{Henrik Ronellenfitsch}
\affiliation{Physics of Biological Organization, Max Planck Institute for Dynamics and Self-Organization (MPIDS), 37077 G\"ottingen, Germany}
\affiliation{Department of Physics and Astronomy, University of Pennsylvania, Philadelphia PA 19104, USA}

\author{Xiaozhu Zhang}
\affiliation{Network Dynamics, Max Planck Institute for Dynamics and Self-Organization (MPIDS), 37077 G\"ottingen, Germany}

\author{Sarah Hallerberg}
\affiliation{Network Dynamics, Max Planck Institute for Dynamics and Self-Organization (MPIDS), 37077 G\"ottingen, Germany}
\affiliation{Fakult\"at Technik und Informatik, Hochschule f\"ur Angewandte Wissenschaften Hamburg,
20099 Hamburg, Germany}

\author{Dirk Witthaut}
\affiliation{Forschungszentrum J\"ulich, Institute for Energy and Climate Research -
	Systems Analysis and Technology Evaluation (IEK-STE),  52428 J\"ulich, Germany}
\affiliation{Institute for Theoretical Physics, University of Cologne, 
		50937 K\"oln, Germany}

\author{Marc Timme}
\affiliation{Network Dynamics, Max Planck Institute for Dynamics and Self-Organization (MPIDS), 37077 G\"ottingen, Germany}
\affiliation{Faculty of Physics, Georg August University G\"ottingen, 37073 G\"ottingen,
  Germany}
\affiliation{Department of Physics, University of Darmstadt, 64289 Darmstadt, Germany}
   
\date{\today }

\begin{abstract}
We introduce the concept of \textit{network susceptibilities} quantifying the 
response of the collective dynamics of a network to small parameter changes.   
We distinguish two types of susceptibilities: \textit{vertex susceptibilities} and \textit{edge 
susceptibilities}, measuring the responses due to changes in the properties of 
units and their interactions, respectively.   We derive explicit forms of 
network susceptibilities for oscillator networks close to steady states and 
offer example applications for Kuramoto-type phase-oscillator models, power 
grid models and generic flow models.   Focusing on the role of the network 
topology implies that these ideas can be easily generalized to other types of 
networks, in particular those characterizing flow, transport, or spreading 
phenomena.   The concept of network susceptibilities is broadly applicable and 
may straightforwardly be transferred to all settings where networks responses 
of the collective dynamics to topological changes are essential.  

\end{abstract}

\maketitle

% --- Content --------------------------------------------------------------------

\section{Introduction}
Susceptibility constitutes a key concept in physics, from statistical 
mechanics to condensed
matter theory and experiments.   In these fields, susceptibility quantifies 
the change of a systems' state, typically measured by order parameters, in 
response to a change in some external field.  
In simple settings, susceptibility is well approximated by linear response 
theory and one global order parameter changes in response.   Generally, there 
can be many order parameters,
as for instance the a site-dependent average spin in the theory of magnetism.  
While ideal solids are organized in the form of perfectly
periodic crystals with, e.g., nearest neighbor interactions, many natural 
and engineered complex systems are organized in networks with a
rich variety of their underlying interaction topologies \cite{Stro01,Newm10}.
The susceptibility
of such a networked system, i.e. its response to changes in their 
parameters, is thus essentially determined by
their topology.   Furthermore, unlike in periodic systems the response depends
crucially on the location of the perturbation.   Given that there are 
different types of local properties that may change, it is not yet 
clear how to appropriately define susceptibilities in a networked system and 
consequentially what such susceptibilities would tell us about the collective 
dynamics.

In this article we introduce two types of susceptibilities in network 
dynamical systems.  
Focusing on changes to steady state operating points, we 
first systematically study the impact of
small local perturbations of single units and effective interactions in
networks. As a key class of network dynamics, we analyze the susceptibilities of 
oscillator networks describing the dynamics of various natural and
man-made systems.   We define both \textit{vertex susceptibilities} and 
\textit{edge susceptibilities} to qualitatively and quantitatively 
distinguish the responses to changes of single unit and single interaction 
properties, respectively.   In particular, we reveal how the interaction 
topology of the network jointly with the type and location of the
perturbation relative to the response location determine the response 
strength.   These susceptibilities are shown to be related to, but not equal 
to, established measures of network centrality. Several applications,
in particular to Kuramoto phase oscillator and power grid networks, are
discussed.   We specifically identify certain instances of vertex
susceptibilities for electric power grid models as power transfer
distribution factors known in electric engineering. Network susceptibilities
are readily generalizable to all kinds of supply and transport networks as
well as network dynamical systems whose dynamics exhibits a
standard flow structure.

\section{Network Susceptibilities}
A continuous time network dynamical system can be described by the equations 
of motion of $N$ variables (the ``vertices'')
\begin{align}
    \label{eq:eom-general}
    \frac{d x_i}{dt} & = F_i(x_1, x_2, \cdots, x_N; p_1, p_2, \cdots, p_M),
\end{align}
where $ p_1,p_2, \ldots, p_M$ are tunable.
The network interactions (the ``edges'') are defined by which variables $x_j$ appear
in the equation of motion of $x_i$. Now we define network 
susceptibilities in the following.  
\begin{definition}
    Let $\vec{x^*} = (x_1^*, x_2^*,\cdots, x_N^*)$ be a steady state for a 
    network dynamical system defined by \eqref{eq:eom-general}. Suppose that 
    on applying a small perturbation to one of the network parameters $p_k$:
\begin{align}
    \label{}
    p_k \to p_k + \varepsilon,
\end{align}
    the fixed point changes by a certain amount
    \begin{align}
        \vec{x^*} & \to \vec{x^{*'}}(\varepsilon).
    \end{align}
    Then the network susceptibility due to parameter $p_k$ is defined as
\begin{align}
    \label{}
    \chi_{(p_k)\to j} & = \lim_{\varepsilon\to 0}\frac{x_j^{*'}(\varepsilon)-x^*}{\varepsilon}.
\end{align}
\end{definition}
We note that this definition can easily be extended to dynamics with other invariant sets 
(e.g.  limit cycles) instead of fixed points, and also to stochastic dynamics.   

\section{Dynamics of oscillator networks}

As a cornerstone example we analyze the susceptibility of a network
of coupled oscillators. The celebrated Kuramoto model \cite{Kura75}
characterizes the collective dynamics of a variety of dynamical systems 
ranging from chemical reactions  \cite{Kura84} and  neural networks \cite{Somp90}
to coupled Josephson junctions \cite{Wies96}, laser arrays \cite{Vlad03}
and optomechanical systems \cite{Hein11}.  In the Kuramoto model, 
$N$ phase oscillators are coupled via their phase differences. The rate 
of change of each phase $\phi_j$ is given by 
\be
  \frac{d \phi_j}{dt} = \omega_j + 
     \sum_{\ell = 1}^{N} K_{j\ell}  \sin(\phi_\ell - \phi_j),
     \label{eqn:kuramoto-intro}
\ee
where $\omega_j$ is the intrinsic frequency of the $j$th 
oscillator, $j \in \{1,\ldots,N\}$,  and $K_{j \ell} = K_{\ell j}$ 
denotes the coupling strength of two oscillators
$j$ and $\ell$. 

A similar model describes the frequency dynamics of complex 
power grids and has gained a strong interest recently
\cite{12powergrid,12braess,Dorf13,Mott13,13powerlong,Nish15}.
The model describes the dynamics of rotating synchronous generators 
and motors, representing power plants and consumers, respectively. 
Each machine is characterized by the power it generates ($P_j > 0$) 
or consumes ($P_j < 0$) and rotates with a frequency close to 
the grid's reference frequency $\Omega$ of $2\pi \times 50/60$ Hz, such 
that its phase is written as $\theta_j(t) = \Omega t + \phi_j(t)$. 
The dynamics of the phases is given by the \emph{swing equation}
\cite{Mach08,Fila08}
\be
   M_j \frac{d^2 \phi_j}{dt^2} + D_j \frac{d \phi_j}{dt}
    = P_j  + \sum_{\ell = 1}^{N} K_{j\ell}  \sin(\phi_\ell - \phi_j), %\sum_{k=1}^N F_{jk},
   \label{eqn:eom-theta}
\ee
where $M_j$ is proportional to the moment of inertia and $D_j$ is proportional 
to the damping torque
of the respective synchronous machine. This `oscillator model' assumes
that all consumers can be described as synchronous motors with a non-vanishing
inertia $M_j$ (It should be noted that since the oscillator model is valid 
only for the high voltage transmission grid, the consumers do not represent 
individual electrical devices in each household, but rather whole cities or 
neighborhoods). In the `structure-preserving model' used in electric power 
engineering
\cite{Berg81} one assumes different consumers.
In contrast to a synchronous machine this type of consumer cannot store any kinetic 
energy, such that the inertia vanishes. Hence, the equations of motion of the
structure-preserving model are still given by (\ref{eqn:eom-theta}),
but with $M_j=0$. 
In the oscillator model as well as the structure-preserving model the power flow from machine 
$k$ to machine $j$ is given by
\be
   F_{jk} = K_{jk} \sin(\phi_k - \phi_j),
    \label{eqn:def-flow}
\ee
where $K_{jk}$ is the maximum transmission capacity which is proportional 
to the susceptance of the respective transmission line. 
The relative load of the transmission line is defined as
\be
   L_{jk} := \frac{F_{jk}}{K_{jk}} = \sin(\phi_k - \phi_j).
   \label{eqn:def-load}
\ee

The two models admit different forms of synchrony. The Kuramoto model was initially introduced to study the emergence of partial synchronization when the coupling of the oscillators in increased \cite{Kura75}. A power grid must be operated in a state of perfect synchronization: All phase differences $\phi_k - \phi_j$ must be constant in time to enable a steady power flow (\ref{eqn:def-flow}). In this article we analyze how such a phase-locked state responds to a local change in the network, and in particular how this change depends on
the topology of the network. 

Transforming to a co-rotating frame, the phase-locked states are then just the steady states of Eq.~(\ref{eqn:kuramoto-intro})  or Eq.~(\ref{eqn:eom-theta}), respectively, which are determined by the algebraic equation
\be
     0 = P_j +  \sum_{\ell = 1}^{N} K_{j \ell}  \sin(\phi_\ell - \phi_j),
  \label{eqn:def-steady}
\ee
such that we can treat the Kuramoto model and the power grid
model on the same footing. However, the perspective of a flow 
network is particularly helpful in understanding the mathematical 
results introduced below. We note that the steady states do not depend
on the mechanical properties of the individual machines, i.e. the moments of inertia $M_j$ 
and the damping coefficients $D_j$.

\section{Linear response theory and network susceptibilities}

In a complex network there are two general scenarios for a microscopic
change of the dynamical system: (1) the modification of an edge weight 
(signifying e.g. a electrical transmission line capacity)
or (2) the modification of a vertex property (e.g. the power generation of a 
power plant) of the system.  In the following, we introduce a
linear response theory for both scenarios.

\subsection{Perturbation at a single edge}
\label{sec:pert-edge}

In the first scenario we consider the 
coupling matrix $K_{ij}$ being perturbed slightly to yield the new perturbed 
matrix $K_{ij}'$, which differs from $K_{ij}$ only at a single edge $(s, t)$:
\bea
K'_{ij} &=& K_{ij} + \kappa_{ij}\\
\kappa_{ij} &=& \left\{
   \begin{array}{l c l}
    \kappa & \;  \mbox{ for } \; &  (i,j) = (s,t) \, \mbox{and} \, (i,j) = (t,s) \\
    0 & & \mbox{all other edges.} \\
   \end{array} \right.         
   \label{eqn:allkappa}
\eea
This perturbation causes the steady state phases 
of the network to change from $\phi_j$ to $\phi'_j$.   The new steady state 
equation \eqref{eqn:def-steady} now reads
\bea
\quad  0 & = & P_ j + \sum_{i=1}^N K'_{ij} \sin(\phi'_i - \phi'_j), \\
         & \forall & j \in \left\{ 1, \ldots, N \right\}. 
  \label{eqn:def-steady-perturbed}
\eea

In the following we calculate
this perturbation within a linear response theory. We note that the 
steady state is defined only up to a global phase shift. Throughout this
article  we fix this phase such that $\sum \nolimits_j \phi_j = 0$.

We expand the steady state condition \eqref{eqn:def-steady-perturbed} to
leading order in  $\kappa_{ij}$ and 
\be
\xi_j := \phi_j' - \phi_j,
\ee 
and subtract \eqref{eqn:def-steady}, to obtain
\bea
   0 &=& \sum_{i=1}^N \kappa_{ij} \sin(\phi_i-\phi_j) + 
        \sum_{i=1}^N K_{ij}\cos(\phi_i-\phi_j) (\xi_i-\xi_j) \nn \\
     &=& \kappa L_{st} (\delta_{js} - \delta_{jt})
             - \sum_{i=1}^N   A_{ji} \xi_i 
   \label{eqn:steady2}
\eea
for all $j=1,\ldots,N$ using the Kronecker symbol $\delta$.
In the last step we have used the definition of the flow (\ref{eqn:def-flow}),
the definition of relative load \eqref{eqn:def-load} and the perturbation
matrix (\ref{eqn:allkappa}). Furthermore, we have introduced the matrix
\bea
    A_{ij} &:=& \left\{
    \begin{array}{l c l}  
       - \widetilde K_{ij}  & \;  \mbox{for} \; & i \neq j \\
      + \sum_\ell  \widetilde K_{\ell j}   & & i=j \\
    \end{array} \right.
    \qquad  \mbox{where} \nn \\
     \widetilde K_{ij} &:=&  K_{ij} \cos(\phi_i - \phi_j).
    \label{eqn:def-laplace}
\eea
In a short-hand vectorial notation,  Eq.~(\ref{eqn:steady2}) then reads
\be
  A \vec \xi = \kappa L_{st}   \vec q_{(st)},
  \label{eqn:Kxiq}
\ee
using the vector $\vec q_{(st)} \in \mathbb{R}^N$ with the components 
\be
    \vec q_{(st),j}  = (\delta_{s,j} - \delta_{t,j}).
   \label{eqn:def-qst}
\ee

We note that the matrix $A$ is singular such that it cannot
be inverted. However, the vector $\vec q$ is orthogonal
to the kernel of $A$ which is spanned by the vector
$(1,1,\ldots,1)^t$ such that this is no problem. In order
to formally solve Eq.~(\ref{eqn:Kxiq}) we can thus use
the Moore-Penrose pseudoinverse of $A$, which we will
call $T := A^+$ in the following. Thus we find
\be
    \vec \xi =  \kappa L_{st} \;   T \vec q_{st}.
   \label{eqn:xi-edge}
\ee

The perturbed flow (\ref{eqn:def-flow}) over an edge $(i,j)$ is 
then given by
\bea
   F'_{ij} &=& (K_{ij} + \kappa_{ij}) \sin( \phi_j - \phi_i + \xi_j - \xi_i ) \nn \\
      &=& K_{ij} \sin( \phi_j - \phi_i)
             + \kappa_{ij} \sin( \phi_j - \phi_i) \nn \\
         && \quad  + K_{ij} \cos( \phi_j - \phi_i) (\xi_j - \xi_i ) 
\eea
up to first order in $\kappa$ and $\vec \xi$.
Using Eq.~(\ref{eqn:xi-edge}), this result reads
\bea
    F'_{ij} = F_{ij} + && \kappa L _{st} \big[
          (\delta_{is} \delta_{jt} - \delta_{js} \delta_{it}) \nn \\
                &&    + \widetilde K_{ij} (T_{js} - T_{jt} - T_{is} + T_{it})     \big].
   \label{eqn:flow-edge}
\eea

\subsection{Edge susceptibilities}
\label{sec:def-edge-chi}

Depending on the application we want to measure different
effects caused by the perturbation at the edge $(s,t)$. First we
quantify how much the phase of a single oscillator $j$ is affected by
the edge-to-vertex susceptibility, using \eqref{eqn:xi-edge}
\bea
   \chi_{(st) \rightarrow j} &:=& \lim_{\kappa \rightarrow 0}
        \frac{\phi_j' - \phi_j}{\kappa}   \nn \\
    &=& L_{st} ( T_{js} - T_{jt}).
\eea
To measure the change of the oscillator state on a global scale in response to 
perturbation at a single edge, we
define the global edge susceptibility as the norm of the local susceptibilities
\bea
\label{eq:global-edge-susc}
     \chi_{(st)}^2 &:=& \lim_{\kappa \rightarrow 0} 
       \frac{\sum_j |\phi_j' - \phi_j|^2}{\kappa^2} \nn \\
    &=& \sum_{j=1}^N   \chi_{(st) \rightarrow j}^2 \nn \\
    &=&  L_{st}^2  \sum_{j=1}^N  ( T_{js} - T_{jt})^2 \, . 
\eea

For applications to flow networks, such as the power grid model
(\ref{eqn:eom-theta}), we are especially interested in how the
\emph{flows} change as this determines the stability of the grid. 
In particular, stability can be lost when a single edge becomes 
overloaded. Thus, we define the edge-to-edge susceptibility as 
the change of flow at another edge
\begin{equation}
 \eta_{(st) \rightarrow (ij)} := \lim_{\kappa \rightarrow 0} 
        \frac{F_{ij}' - F_{ij}}{\kappa} \, .
  \label{eqn:def-eta}
\end{equation}
Using Eq.~(\ref{eqn:flow-edge}) this relation reads
\bea
   \eta_{(st) \rightarrow (ij)} = L_{st} \big[ &&
    (\delta_{is} \delta_{jt} - \delta_{js} \delta_{it}) \nn \\
       && +   \widetilde K_{ij} (T_{js} - T_{jt} - T_{is} + T_{it}) \big] .
    \label{eqn:eta-ee-T}
\eea

We conclude that the effects of a perturbation at a single edge
$(s,t)$ as measured by the susceptibilities defined above are 
proportional to the \emph{load} of edge $L_{st}$. Furthermore,
the susceptibilities are essentially given by the matrix $T$, the pseudoinverse
of $A$. The properties of these matrices will be analyzed in detail
in the following sections.

\subsection{Perturbation at a single vertex}

The above calculations can be readily generalized to analyze the 
change of the steady state in response to a local perturbation of
a single vertex property. 
To this end we consider a change of the power injected at a single vertex $s$.
However, a steady state of  Eq.~(\ref{eqn:eom-theta})
exists only if the power is balanced such that we consider
a small perturbation of the power vector of the form
\be
    P_j' = P_j + p \, (\delta_{j,s} - 1/N).
   \label{def:pert-vertex}
\ee
Expanding the definition of a steady state to leading order in $p$ 
and $\xi_j := \phi_j' - \phi_j$ then yields
\bea
   0 &=& p \, (\delta_{j,s} - 1/N) 
             +  \sum_{i=1}^N \widetilde K_{ij} (\xi_i-\xi_j)   \nn \\
     &=& p \, (\delta_{j,s} - 1/N) 
             +  \sum_{i=1}^N A_{ij} \xi_i  .
   \label{eqn:steady3}
\eea
Solving this equation for the changes $\vec \xi$ yields
\be
   \vec \xi = p \;  T \vec r_s
\ee
with the vector $\vec r_s \in \mathbb{R}^N$ whose components are given by
\be
    r_{s,j} = \delta_{s,j} \, .
\ee

\subsection{Vertex susceptibilities}
\label{sec:vertex-sus}

In analogy to the case of a perturbed edge discussed in
section (\ref{sec:def-edge-chi})
we define the vertex-to-vertex susceptibility as
\bea
    \chi_{s \rightarrow j} &:=& 
         \lim_{p \rightarrow 0} \frac{\phi_j' - \phi_j}{p} \nn \\
        &=& T_{js},\
    \label{eqn:xi-vertex}
\eea
the global vertex susceptibility as
\bea
    \chi_{s}^2 &:=& \sum_{j=1}^N \chi_{s \rightarrow j}^2  \nn \\
        &=& \sum_{j=1}^N T_{js}^2,
\eea
and the vertex-to-edge susceptibility as
\bea
    \eta_{s \rightarrow (ij)} &:=& \lim_{p \rightarrow 0} 
              \frac{F_{ij}' - F_{ij}}{p}   \nn \\
     &=& \widetilde K_{ij} ( T_{js} - T_{is} ).
\eea

We note that measures similar to the vertex-to-vertex susceptibility $\chi_{s \rightarrow j}$ are used in electric power engineering where they are called power transfer distribution factors \cite{Wood13,16ptdf}. In this context one generally uses a fixed reference or slack node which absorbs the power change $p$, such that Eq.~(\ref{def:pert-vertex}) is modified to
\be
    P_j' = P_j + p \, (\delta_{j,s} - \delta_{j,\rm slack}).
\ee

\subsection{Properties of the matrix $A$}

We have shown that the response of a network to a local
perturbation is essentially given by the matrix $T$, which is 
the Moore-Penrose pseudoinverse of the matrix $A$ defined 
in Eq.~(\ref{eqn:def-laplace}). Before we discuss the potential 
applications of the network susceptibilities we thus have a closer
look at the properties of the matrix $A$.

The matrix $A$ encodes the dynamical stability and synchrony of steady states
\cite{14bifurcation,Skar14}. A steady state of the Kuramoto model or
the power grid model defined by Eq.~(\ref{eqn:eom-theta}) is dynamically stable
if and only if $A$ is positive semi-definite, i.e.~all its eigenvalues 
$a_j,  1 \leq j \leq N$ are non-negative. For sake of simplicity we fix the ordering
of the eigenvalues such that $0 = a_1 \le a_2\le a_3\le\cdots a_N$. We have to take 
into account that $A$ always has one eigenvalue $a_1 = 0$.   The corresponding 
eigenvector is $(1,1,\cdots,1)$; signifying that a small perturbation
that is exactly the same in all phase angles is neutrally stable.  However, this is 
merely due to the steady state itself being arbitrary up to a constant global 
phase shift. Stable steady states can emerge or disappear
when a system parameter is varied through an (inverse) saddle
node bifurcation at which one eigenvalue vanishes, $a_2 \rightarrow 0$.

In particular, $A$ is positive semi-definite if the relation 
$\cos(\phi_i - \phi_j) > 0$ holds for all edges $(i,j)$ of the 
network and the network is globally connnected. Stable steady states which do 
not satisfy this relation typically exist only at the edge of the stable 
parameter region
\cite{14bifurcation}. We can thus assume that 
during normal operation we always have  $\cos(\phi_i - \phi_j) \ge 0$
for all edges such that we can use the following relations:
\bea
    \cos(\phi_i-\phi_j) &=& \sqrt{1 - \sin(\phi_i-\phi_j)^2 }  \ge  0\nn \\
   \Rightarrow  \widetilde K_{ij} =  K_{ij}\cos(\phi_i-\phi_j) &=& \sqrt{K_{ij}^2 - F_{ij}^2}.
\eea
The expression $\widetilde K_{ij}$ can be understood as the free capacity of an edge $(ij)$, which can be used to respond to the
perturbation and is thus refereed to as the \emph{responsive capacity}.

For normal operation, $\cos(\phi_i - \phi_j) \ge 0$ for all edges $(i,j)$,  the non-diagonal entries of the matrix $A$ are all non-positive such that $A$ is a \emph{Laplacian matrix}
for which many properties are known \cite{Newm10}.
In particular, the eigenvalues of a Laplacian matrix satisfy 
$0 = a_1 \le a_2 \le \cdots \le a_N$, where $a_2$ is 
an algebraic measure for the \emph{connectivity} of the
underlying network  \cite{Fied73, Fort10}.

\section{Susceptibility and connectivity}

\subsection{Scaling properties of network susceptibilities}
\label{sec:globalsus}

The susceptibilities are especially large in the limit of a weakly connected 
network. For a power grid this corresponds to the scenario of high loads when
the responsive capacities $\widetilde K_{ij}$ become small. In the following 
we analyze this case in detail for a perturbation at a single 
edge $(s,t)$ (cf.~Eq.~(\ref{eqn:allkappa})). The case of a 
vertex perturbation is discussed briefly at the end of this section.

\begin{figure*}[tb]
\centering
\includegraphics[width=4.5cm]{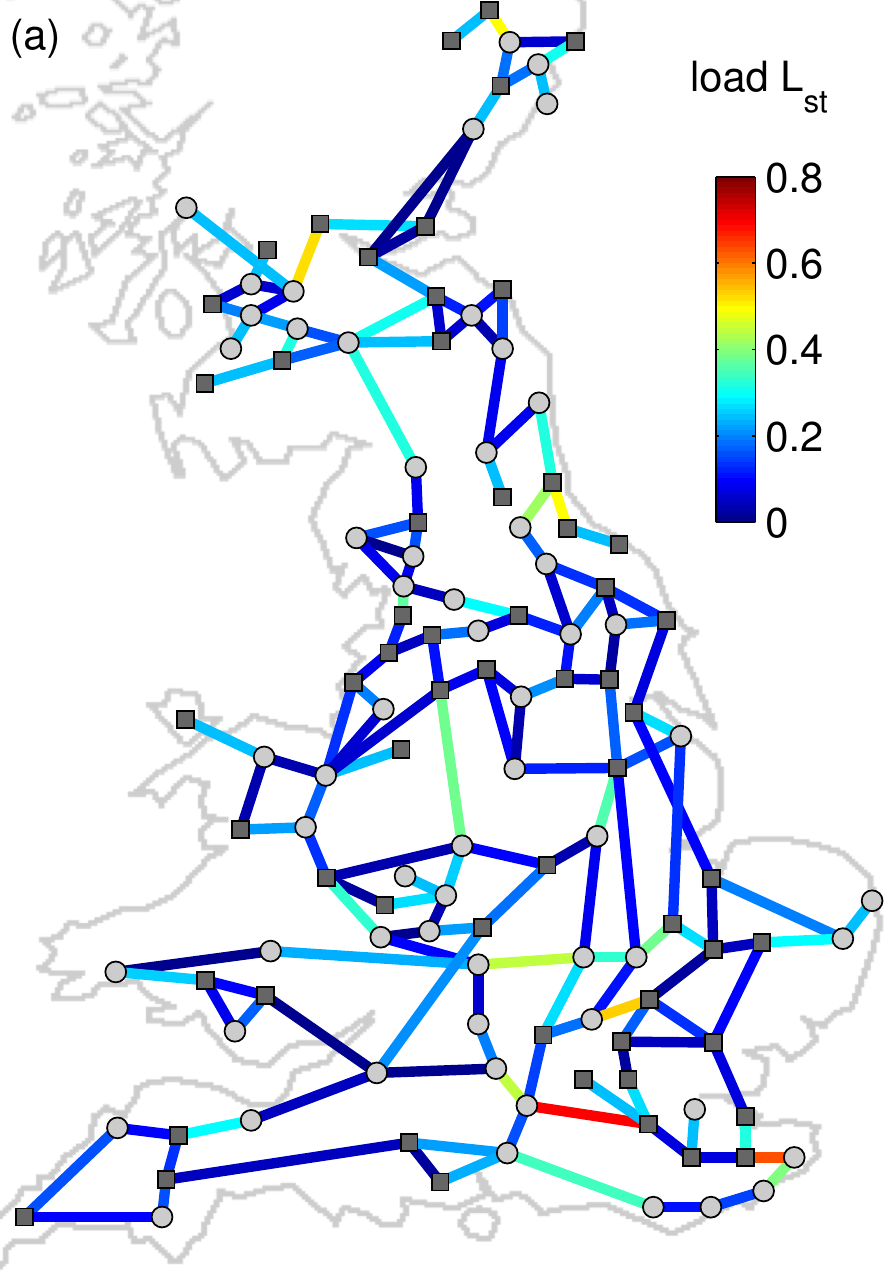}
\hspace{5mm}
\includegraphics[width=4.5cm]{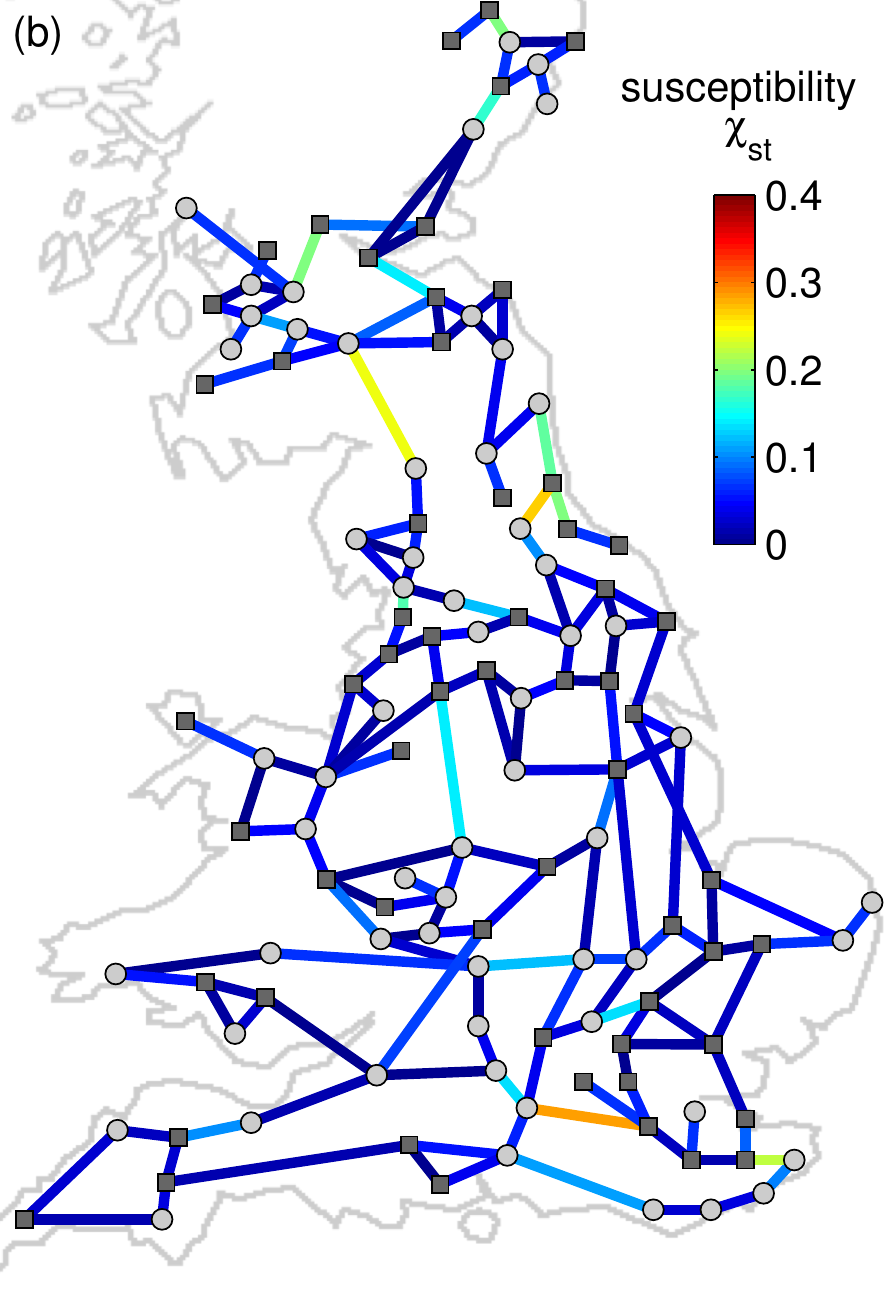}
\hspace{5mm}
\begin{minipage}[b]{6.6cm}
\includegraphics[width=6.6cm]{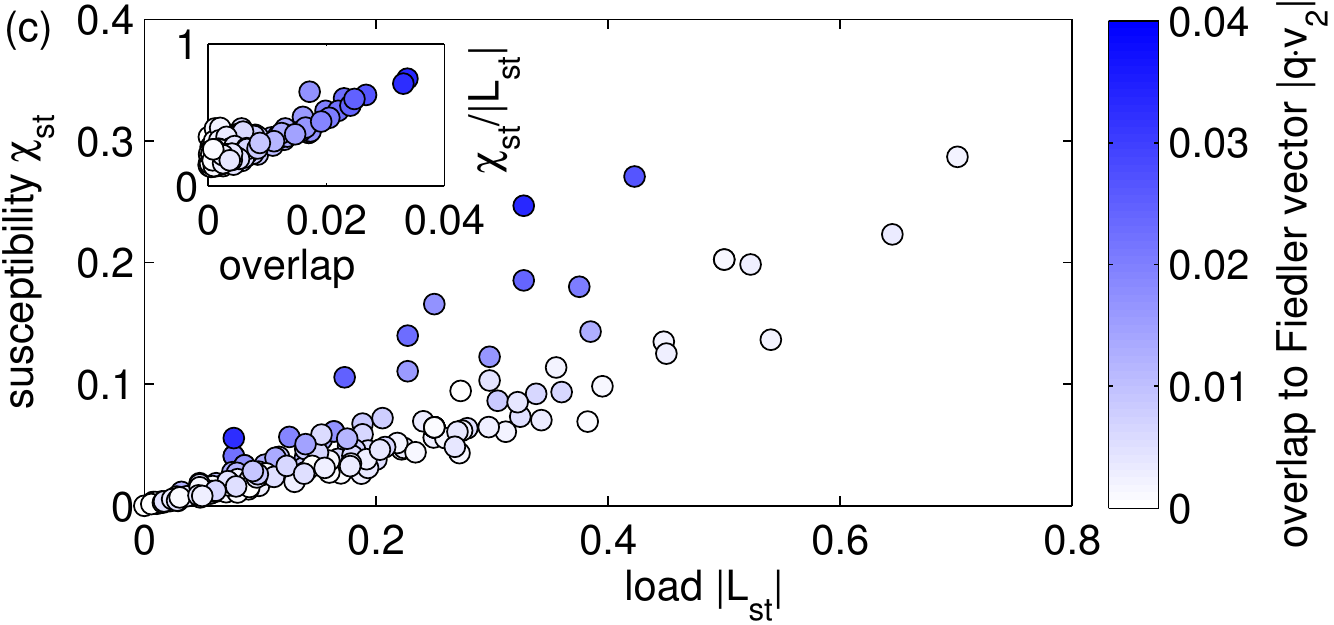}
\includegraphics[width=6.6cm]{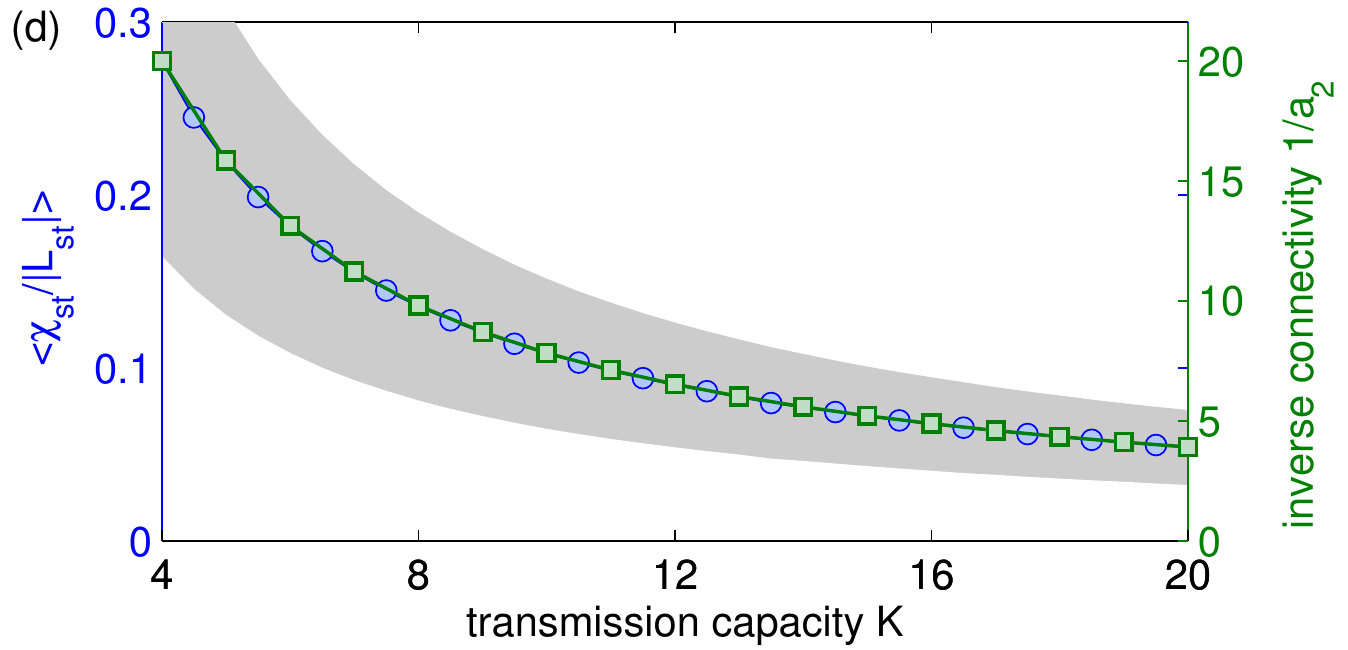}
\end{minipage}
\caption{
\label{fig:globalsus}
The global edge susceptibility $\chi_{(st)}$ in a model power grid.
(a) Coarse-grained topology of the British high-voltage power
transmission grid \cite{Simo08,12powergrid}. We randomly choose 60 nodes to be generators
with $P_j = +1$ ($\square$) and 60 nodes to be consumers with 
$P_j = -1$ ($\circ$). The transmission capacity of all edges is given by $K = 4$
in arbitrary units.
The color map shows the load $|L_{st}|$ of each edge.
(b) Color map plot of the global edge susceptibility $\chi_{st}$. 
(c) For a given network, the susceptibility is approximately
proportional to the load of the edge $|L_{st}|$. 
It is increased if the edge $(s,t)$ couples two weakly connected
components of the responsive capacity graph $\widetilde K$, indicated
by a large overlap with the Fiedler vector $|\vec q_{st} \cdot  \vec v_2|$
(shown as a color code and in the inset).
(d) On a global scale, the average susceptibility is proportional
to the inverse algebraic connectivity $1/a_2$. The plot shows
$1/a_2$ ($\square$, right scale) and the ratio $\chi_{st}/|L_{st}|$ averaged 
over all edges ($\circ$, left scale) as a function of the 
transmission capacity $K$. The shading shows the 
standard deviation of $\chi_{st}/|L_{st}|$.
}
\end{figure*}

Throughout this section we assume the case of 'normal' operation, i.e. we
assume that $\widetilde K_{ij} \ge 0$ for all edges $(i,j)$. Then the matrix
$A$ is a Laplacian matrix with eigenvalues $0 = a_1 \le a_2 \le \cdots \le a_N$
and the associated eigenvectors $\vec v_n$. We can then formally
solve Eq.~(\ref{eqn:Kxiq}) for $\vec \xi$ with the result
\bea
     \vec \xi &=& \kappa L_{st} \sum_{n=2}^N  \frac{1}{a_n}     
          ( \vec v_n \cdot \vec q_{st} )   \vec v_n .
    \label{eqn:xi-eigenvectors}
\eea
The term $n=1$ does not contribute since we have fixed the global
phase such that $\sum \nolimits_j \xi_j = 0$.
This expression shows four important properties of the
network susceptibility:

(1) The response $\vec \xi$ and thus also the edge susceptibilities scale with the \emph{load} of the 
perturbed edge $L_{st} = F_{st}/K_{st}$. 
For a complete breakdown of an edge $(s,t)$, we have 
$\kappa = - K_{st}$ such that $\vec \xi$ scales with the flow
$F_{st}$ of the defective edge. The scenario of a complete breakdown
is further discussed in Sec.~\ref{sec:crit-lines}.
   
(2) The prefactors $1/a_n$ decrease with $n$.
In particular for a weakly connected network the algebraic connectivity 
$a_2$ becomes very small \cite{Fied73, Newm10, Fort10}, such that the 
term $n=2$ dominates the sum. Then the susceptibility of all edges in 
the network scale inversely with the algebraic connectivity $a_2$.
This proves our claim that the susceptibility is large if the network defined 
by the responsive  capacities $\widetilde K_{ij}$ is weakly connected.

(3) For a weakly connected network, the edge susceptibility scales
with the overlap $|\vec v_2 \cdot \vec q_{st}|$, where $\vec v_2$
is the so-called Fiedler vector. This overlap can be interpreted
as a measure of the \emph{local algebraic connectivity} of the 
nodes $s$ and $t$. To see this note that the Fiedler vector
can be used to partition a graph into two weakly connected
parts \cite{Newm10,Fort10}. The overlap with the vector $\vec q_{st}$ 
is largest if the two nodes $s$ and $t$ are in different parts 
and thus weakly connected.

(4) In the limit of a disconnected network  the response $\vec \xi$ to
a perturbation at the edge $(s,t)$ diverges if the edge links the
weakly-connected components. If the perturbation occurs within
one component, then the response remains finite. This will be shown in detail
in the following section.

(5) The global edge susceptibility defined in Eq.~(\ref{eq:global-edge-susc}) can be expressed as
\be
    \chi_{(st)}^2 = L_{st}^2 \sum_{n=2}^N \frac{(\vec v_n \cdot \vec q_{st})}{a_n^2},
\ee
where we have used Eq.~(\ref{eqn:xi-eigenvectors}) for the phase response.
This quantity measures the average phase response to the perturbation of a single edge $(s,t)$.
An example is shown in Fig.~\ref{fig:globalsus} for a synthetic power grid model based on the topology of the British high-voltage grid. One observes that the global susceptibility of an edge $(s,t)$ is essentially determined by the load $L_{st}$, the connectivity of the network and the location of the edge within the network. Edges are highly susceptible if they are heavily loaded or connect two components of the grid. In the shown example we observe two highly susceptible edges connecting the northern part to the rest of the grid.
Averaging the global susceptibilities $\chi_{st}$ over all edges $(s,t)$ in the network, we find an almost perfect proportionality with the inverse algebraic connectivity $1/a_2$. If the transmission capacity $K$ of the edges increases, the algebraic connectivity $a_2$ also increases and the grid becomes less susceptible to perturbations.

\subsection{The weakly connected limit}

To obtain a more quantitative 
understanding of the susceptibility in a weakly connected network we 
assume that the network is decomposed into two components 
of size $N_1$ and $N_2 = N - N_1$, respectively. In the limit of complete 
disconnection, the Laplacian matrix also decomposes
\be
   A^{(0)} = \begin{pmatrix} 
               A_1^{(0)}   & 0 \\
               0   &  A_2^{(0)}  \\
     \end{pmatrix}  
\ee
with $A_1^{(0)}  \in \mathbb{R}^{N_1 \times N_1}$ and
$A_2^{(0)}  \in \mathbb{R}^{N_2 \times N_2}$. As usual for
a Laplacian matrix the lowest eigenvalue vanishes, $a_1= 0$,
and the associated eigenvector is given by
\be
   \vec v_1 = \frac{1}{\sqrt{N}}  (1,1,\ldots,1)^T.  
\ee
In the disconnected limit also the second eigenvalue (the algebraic connectivity),  
vanishes, $a_2^{(0)} = 0$. The associated eigenvector,
the Fiedler vector, is given by
\be
   \vec v_2^{(0)} = \frac{1}{\sqrt{N}}  \;
     ( \underbrace{ \sqrt{N_2/N_1},\ldots}_{N_1 \; \mbox{times}} \, ,
      \underbrace{-\sqrt{N_1/N_2} ,\ldots}_{N_2 \; \mbox{times}}  )^T .
\ee
Here and in the following the superscript $(0)$ denotes the limiting
case of a complete disconnection of the network. For 
simplicity we assume that the two components are not further 
disconnected such that $a_3^{(0)} > 0$.

To analyze the case of a weakly-connected network, we consider
a single weak link at position $(c,d)$ between the two components. 
The Laplacian is then given by $A = A^{(0)} + A'$  with
\bea
    && A'_{cd} = A'_{dc} = - k  \nn  \\
    && A'_{cc} = A'_{dd} = + k  
\eea
and $A'_{ij} = 0$ otherwise. The connection strength $k$ of the edge
$(c,d)$ is assumed to be small such that we can calculate the
eigenvalues and eigenvectors using 
Rayleigh-Schr\"odinger perturbation theory
(see, e.g., \cite{Ball98}).
We then find the algebraic connectivity
\be
    a_2 =  k \frac{N_1+N_2}{N_1 N_2} + \OO(k^2)
\ee
and the Fiedler vector
\be
  \vec v_2 =  \vec v_2^{(0)} + k \sqrt{\frac{N_1+N_2}{N_1  N_2}}  
                          \sum_{n = 3}^{N} 
                          \frac{(\vec v_n^{(0)}  \cdot \vec q_{cd})}{a_n^{(0)}} 
                      \vec v_n^{(0)}   + \OO(k^2),
\ee
where $\vec q_{cd}$ is defined as in Eq.~(\ref{eqn:def-qst}).

To calculate the response of the network $\vec \xi$ we need the overlap
of the vector $\vec q_{st}$ (see Eq.~(\ref{eqn:xi-eigenvectors})) with
the eigenvectors of $A$, in particular the overlap with the Fiedler vector. 
The result depends crucially on the location of the perturbed edge 
$(s,t)$. If this edge connects the two components, i.e.~$(s,t) = (c,d)$,
we find
\be
     \vec v_2 \cdot \vec q_{st} =  \sqrt{\frac{N_1 + N_2}{N_1  N_2}} + \OO(k)
\ee
such that the response diverges as $k^{-1}$:
\be
    \vec \xi = \frac{\kappa L_{st}}{k}  \sqrt{\frac{N_1  N_2}{N_1 + N_2}} \vec v_2^{(0)} 
     + \OO(k^0).
\ee
If the edge $(s,t)$ lies within one component, then
\be
     \vec v_2 \cdot \vec q_{s,t} = k  \sqrt{\frac{N_1 + N_2}{N_1  N_2}}
          \sum_{n=3}^N \frac{  (\vec v_n^{(0)} \cdot \vec q_{cd}  )
           (\vec v_n^{(0)} \cdot \vec q_{st}  ) }{a_n^{(0)}}   + \OO(k^2) \nn
\ee
such that the response remains finite in the limit $k\rightarrow 0$:
\bea
    \vec \xi &=& \kappa L_{st} \sqrt{\frac{N_1  N_2}{N_1 + N_2}}  \sum_{n=3}^N 
                \frac{  (\vec v_n^{(0)} \cdot \vec q_{cd}  )
           (\vec v_n^{(0)} \cdot \vec q_{st}  ) }{a_n^{(0)}}  \vec v_2^{(0)} \nn \\
       && + \kappa L_{st} \sum_{n=3}^N 
                \frac{ (\vec v_n^{(0)} \cdot \vec q_{st}  ) }{a_n^{(0)}}  \vec v_n^{(0)}.
\eea

For a perturbation at a single vertex as defined in
Eq.~(\ref{def:pert-vertex}) the response will always diverge in the limit
$k \rightarrow 0$. Assuming w.l.o.g. that the perturbed vertex $s$ is an 
element of the component $1$ we find that
\be
  \vec \xi = \frac{p}{k}  \, \frac{N^2}{N_1 N_2} 
     ( \underbrace{ N_2/N_1,\ldots,N_2/N_1}_{N_1 \; \mbox{times}} \, ,
      \underbrace{-1 ,\ldots,-1}_{N_2 \; \mbox{times}}  )^T \,
\ee  
to leading order.

\section{Applications}
\label{sec:applications}

\subsection{The relation to centralities}
\label{subsec:centrality}
Various centrality measures have been defined to quantify the importance of 
single vertices and edges in complex networks\cite{Newm05}. Centrality measures 
based on current flows \cite{Brandes2005} are heavily used in different areas of network 
science and are directly related to susceptibility measures as defined in the 
present article. To illustrate this, consider a network of ohmic resistors with 
conductances $G_{ij}$.
An electrical current flows through the network with $I^{\rm source}_j$ 
being the current in- or outflow at vertex $j$. The current through a
particular edge $(i,j)$ of the network is given by the voltage drop
across the edge such that
\be
   I_{ji} = G_{ji} (V_j - V_i).
\ee
At each vertex the current is conserved such that Kirchhoff's law
\be
  \sum_{i=1}^N I_{ji} =  \sum_{i=1}^N  G_{ji} (V_j - V_i) = I^{\rm source}_j
\ee
is satisfied for all  $j= 1,\ldots,N$. Defining the Laplacian
matrix of the conductances (the so called nodal conductance matrix)
\bea
    A_{ij} &:=& \left\{
    \begin{array}{l c l}  
       - G_{ij}  & \;  \mbox{for} \; & i \neq j \\
      + \sum_\ell  G_{\ell j}   & & i=j \\
    \end{array} \right.
\eea
and its Moore-Penrose pseudoinverse $T := A^+$, the voltages are given by
\be
   \vec V =  T \vec I^{\rm source} .
\ee
For the definition of centrality measures \cite{Newm05} one considers the 
situation that a unit current flows into the network at a single vertex $s$ and
out at a different vertex $t$. Then we have the voltages
\be
    V_j =  T_{js} - T_{jt}
    \label{eq:volt-from-T}
\ee
and the current flowing over the edge $(i,j)$ is given by
\be
   I_{ji} = G_{ji}  ( T_{js} - T_{jt}   -  T_{is} + T_{it} )  .
\ee
The current flow betweenness centrality of an edge $(i,j)$
is then defined as the absolute current flowing
through the edge averaging over all scenarios 
of the in-/out-flow, i.e. all pairs $(s,t)$ \cite{Newm05}
\be
    b_{(i,j)} := \frac{2}{N(N-1)} \sum_{s<t} 
               G_{ji}  | T_{js} - T_{jt}   -  T_{is} + T_{it} |.
   \label{eqn:def-centrality}
\ee
Correspondingly, the betweenness centrality of a vertex
$j$ is defined as 
\bea
 b_j &:=& \frac{1}{N(N-1)} \sum_{s < t} 
      \sum_{i=1}^N  G_{ji}  | T_{js} - T_{jt}   -  T_{is} + T_{it} | \nn  \\
     &=& \sum_{i=1}^N  \frac{1}{2} \,  b_{(i,j)}.
\eea
We directly see the analogies to the definition of the network 
susceptibilities if we identify the conductance $G_{ij}$ with the 
responsive capacity $\widetilde K_{ij}$. In particular, the
edge betweenness centrality defined in Eq.~(\ref{eqn:def-centrality})
coincides with the average of the normalized edge-to-edge 
susceptibility $\eta_{st \rightarrow ij}/|L_{s,t}|$ except
for a slight difference  in the term $(s,t) = (i,j)$ that vanishes
as $1/N^2$.

However, in this article we generalize the idea of centralities 
based on current flow in several ways. First of all, we consider 
two different scenarios for the in- and outflow of  the network:
First, for the edge susceptibilities we consider an inflow at vertex $s$ and outflow at vertex $t$ 
with strength $L_{st}$ as in \cite{Newm05} 
\be
    I^{\rm source}_j = L_{st} (  \delta_{js} - \delta_{st} ). 
\ee
Second, for the vertex susceptibilities we assume a unit inflow at
vertex $s$ and equal outflow at all other vertices such that
\be
    I^{\rm source}_j =   \delta_{js} - \frac{1}{N} \, .
\ee
Third, we analyze not only the change of the flows as in the
edge-to-edge susceptibilities, but also the change of the 
state variables $\xi_j$ which correspond to the voltages
in the resistor networks. In this sense, the global susceptibilities 
$\chi^2_{(st)}$ and $\chi^2_s$ are given by the \emph{variance} of the 
voltages in the network. Therefore they quantify the \emph{global} 
response of the network  to a local in- or outflow in terms
of the average variation of all voltages.

\subsection{Relation to resistance distances}

In a manner similar to the previous section \ref{subsec:centrality}, the concept of susceptibilities can be understood in terms of \emph{resistance distance}, which is defined as follows. As in the previous section we consider a network of Ohmic resistors with conductances $G_{ij}$ and suppose a unit current enters the node $s$ and exits through node $t$. Then the resistance distance $R_{st}$ is given by the voltage drop between the nodes $s$ and $t$. Using the relation (\ref{eq:volt-from-T}), this yields
\be
   R_{st} = V_s - V_t  = T_{ss} - 2T_{st} + T_{tt}. 
\ee
using the symmetry of the matrix $T$. This relation can be inverted with the result
 \cite{palacios2001closed} 
\begin{align}
    \label{eq:R_T_relation}
    T_{ij} & = -\frac{1}{2}R_{ij} +\frac{1}{2N}\left(R_j^{\rm tot} +R_i^{\rm tot}\right) -\frac{\sum_{i,j}R_{ij}}{N^2}
\end{align}
where we have defined $R_i^{\rm tot} = \sum_j R_{ij}$. 

Substituting Eq.~(\ref{eq:R_T_relation}) into \eqref{eqn:xi-edge} and \eqref{eqn:xi-vertex}, we can express all susceptibilities equivalently in terms of the matrix $T$ or the resistance distances. For the vertex-to-vertex susceptibility we find
\begin{align}
    \chi_{s\rightarrow t} & = -\frac{1}{2}R_{st} + \frac{1}{2N} R_s^{\rm tot} + \frac{1}{2N} R_t^{\rm tot},
\end{align}
and subsequently the global average of susceptibilities take the simple form
\begin{align}
    \sum_{t\neq s} \chi_{s \rightarrow t}  &= \frac{1}{2}\sum_{i,j}G_{ij} - \frac{1}{N} R_s^{\rm tot}.
\end{align}
This relation clearly demonstrates that nodes that are on an average ``close" to the rest of the network (i.e. with high centrality values), tend to have higher global susceptibility. 

In a similar manner, the vertex-to-edge susceptibilities can be expressed as
\begin{align}
    \eta_{s\to(i,j)} &= \widetilde{K}_{ij}\left\{-\frac{1}{2}(R_{si}-R_{sj}) + 
\frac{1}{2N}(R_i^{\rm tot} - R_j^{\rm tot})\right\}
\end{align}
and the edge-to-vertex susceptibility follows an almost identical form, apart from the prefactor:
\begin{align}
    \eta_{(i,j)\to s} &= L_{ij}\left\{-\frac{1}{2}(R_{si}-R_{sj}) + 
\frac{1}{2N}(R_i^{\rm tot} - R_j^{\rm tot})\right\}.
\end{align}
The global edge susceptibilities are given by (derivation in
Appendix \ref{app:global-susc}):
\begin{align}
    \label{eq:global-e2v-susc}
    & \chi^2_{(ij)} =   \\
    & \;  \frac{N L_{st}^2}{4}\left(\frac{1}{N}\sum_s (R_{si} - R_{sj})^2 - \left[\frac{1}{N}\sum_s(R_{si} - R_{sj}) \right]^2\right). \nn 
\end{align}

\subsection{Scaling with distance}
\label{sec-distance}

\begin{figure}
    \includegraphics[width=\columnwidth]{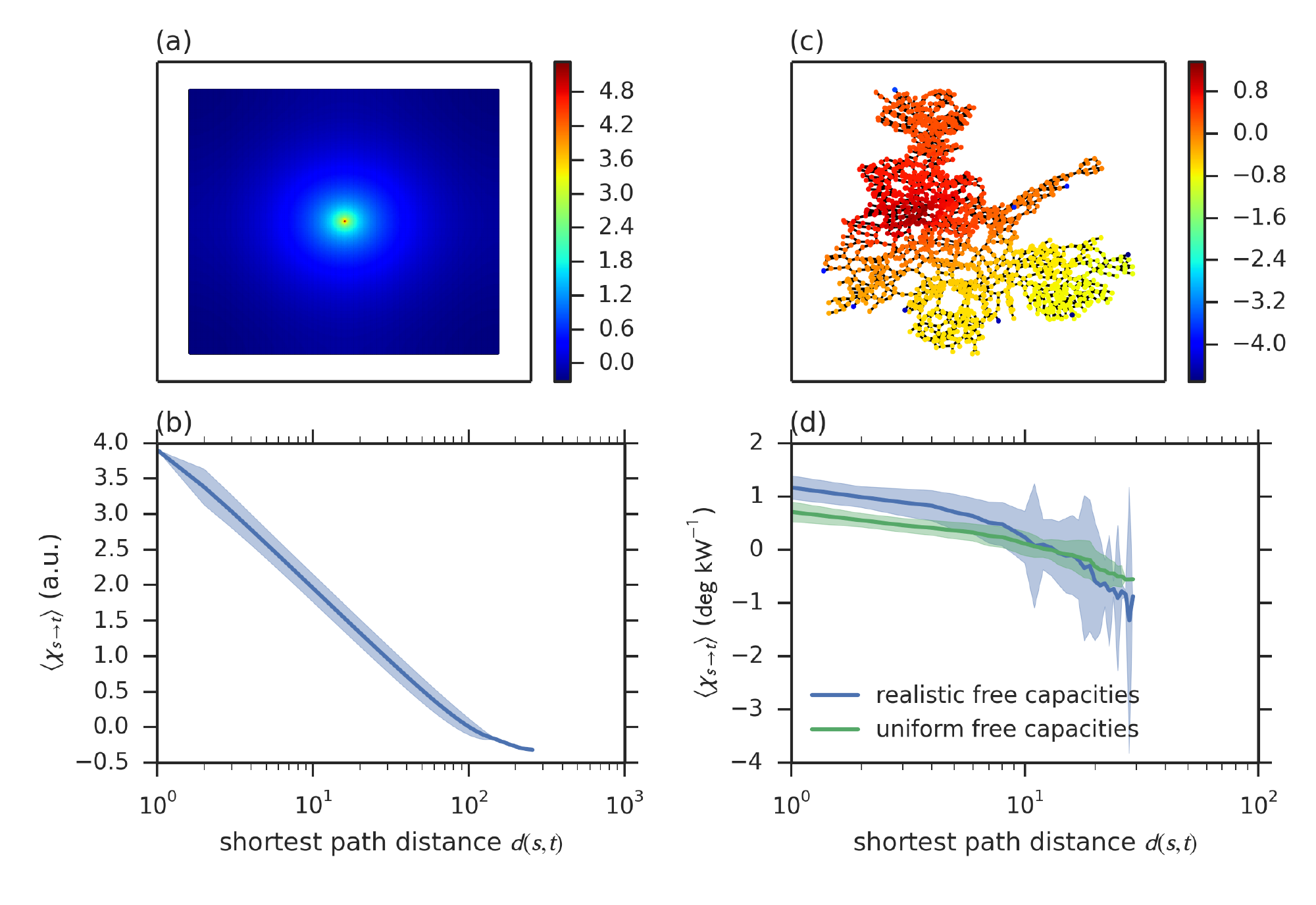}
    \caption{The Vertex-to-vertex susceptibility in uniform and realistic
    network topologies. (a) Color coded plot of $\chi_{s\rightarrow t}$
    in a $256\times 256$ square grid with uniform free capacities, showing 
    logarithmic decay. The central vertex $s$ was perturbed. (b) Decay behavior of the mean
    $\chi_{s\rightarrow t}$ in the same topology as in (a) as a function
    of shortest path distance $d(s,t)$. The shaded region represents
    a 95\% confidence interval.
    We clearly see logarithmic decay.
    (c) Color coded plot of $\chi_{s\rightarrow t}$ in the
    Continental European Transmission Network topology with
    realistic free capacities (taken from \cite{Hutc13}). The vertex
    positions are not realistic.
    One central vertex
    was perturbed. There are several highly susceptible vertices in the
    network periphery (dark blue). (d) Decay behavior of the mean
    $\chi_{s\rightarrow t}$ in the same topology as in (c) as a function
    of $d(s,t)$. We show both realistic free capacities 
    $\tilde K_{ij,\mathrm{real}}$ as well as uniform free capacities
    $\tilde K_{ij,\mathrm{unif}} = 
    \langle\tilde K_{\mathrm{real}}\rangle$ set to the mean realistic
    value. The same vertex as in (c) was perturbed. In the realistic case,
    few highly susceptible vertices in the network periphery lead to a
    high variance at large distances, in contrast to the uniform case.
    \label{fig:vv-susceptibility}
    }
\end{figure}

The effect of a linear perturbation generally decays with distance.
To obtain a better understanding of this decay, we consider a continuum 
version of the linear response theory, concentrating on the vertex-to-vertex susceptibility.
We consider a two-dimensional square lattice with equal weights, as
power grids are naturally embedded into a two-dimensional plane and
most grids can can be assumed to be approximately planar.
In the continuum limit the Laplacian matrix tends to the two-dimensional
Laplace operator and Eq.~(\ref{eqn:steady3}) becomes a Poisson 
equation,
\begin{align}
    \Delta \xi(\vec x) = p \delta(\vec x - \vec x_0),
\end{align}
where $\xi(\vec x)$ is the local response at position $\vec x$ 
(e.g.~the local phase angle), $p$ is the power injection which occurs 
at position $\vec x_0$ and $\Delta$ is the 2D Laplace operator. The solutions to this equation are well known. On an infinite 
two-dimensional domain it is
\begin{align}
    \xi(\vec x) = \frac{p}{2\pi} \log(|\vec x - \vec x_0|) + b,
\end{align}
where $b$ is a constant of integration.
Generally, no unique notion of Euclidean distance between nodes exists 
for networks. The closest analog is the shortest path distance,
denoted by $d(s,t)$ in the following, which is related to the Euclidean 
distance for instance in regular grids.
Figure \ref{fig:vv-susceptibility} (a, b) show the decay behavior
in a uniform square grid, compatible with the continuum results.

Realistic network topologies are more complicated as shown in 
Fig.~\ref{fig:vv-susceptibility} (c, d). We computed the susceptibility of
the Continental European Transmission Network \cite{Hutc13}
to perturbing
one vertex for two cases of free capacities $\tilde K$. First,
we obtained realistic values $\tilde K_{ij,\mathrm{real}}$ from 
\cite{Hutc13}, 
then we considered a uniform model in which all free capacities are
replaced by the average $\tilde K_{ij,\mathrm{unif}} = 
    \langle\tilde K_{\mathrm{real}}\rangle$.
In the vicinity of the perturbation, monotonic decay can be
seen in both cases. However, there exist several vertices in the 
periphery of the network that are much more susceptible than the rest
for realistic free capacities (dark blue in  
Fig.~\ref{fig:vv-susceptibility} (c)). These vertices are highly susceptible
independent of the perturbed vertex.

\begin{figure}
    \includegraphics[width=\columnwidth]{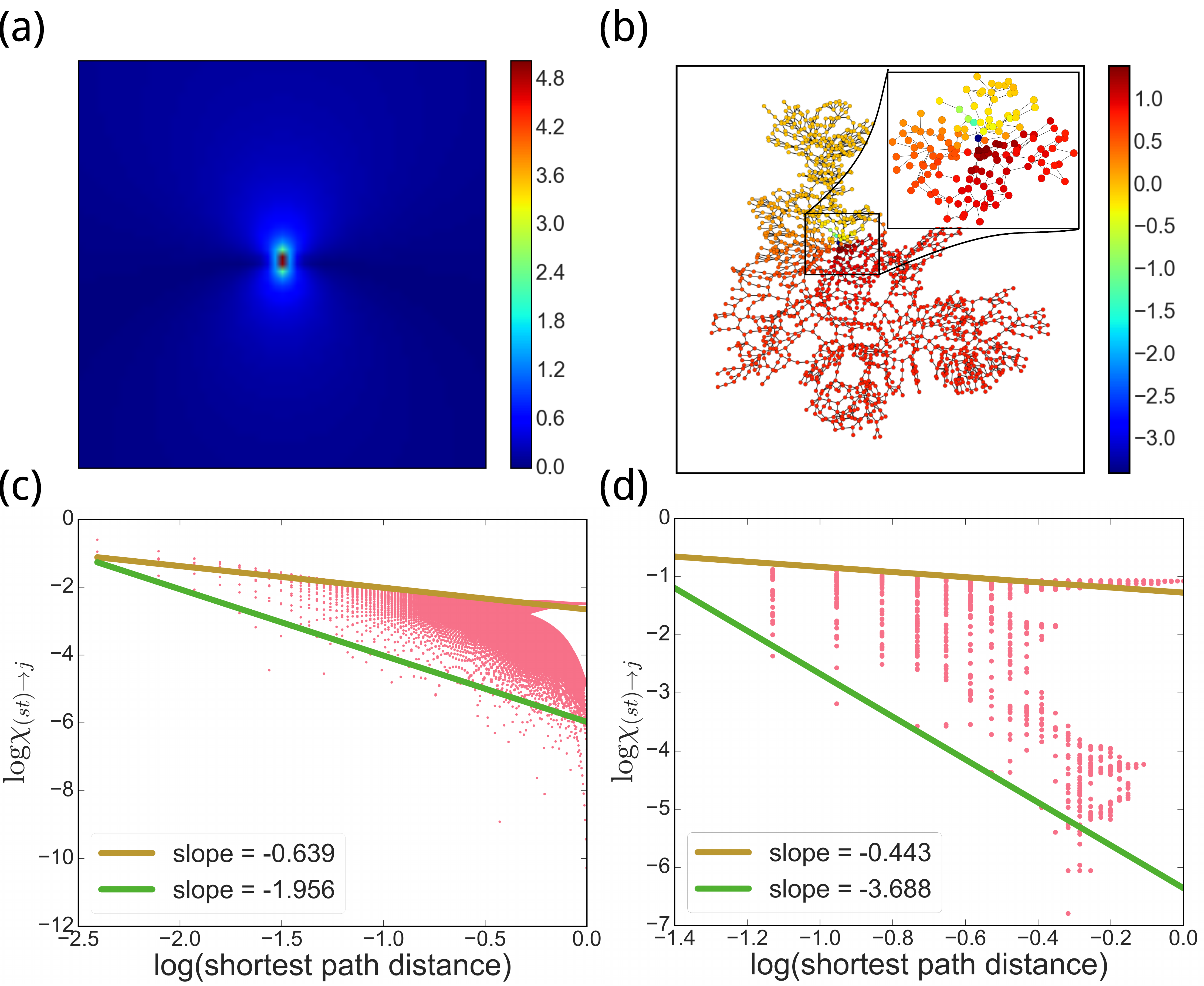}
    \caption{The edge-to-vertex susceptibility in uniform and realistic
        network topologies. (a) Color coded plot of $\chi_{(s,t)\rightarrow j}$
    in a $256\times 256$ square grid with uniform free capacities.
    The central edge ((128,128), (128,129)), where the numbers are integer coordinates, was perturbed.
    (b) Decay behavior of 
    $\chi_{(st)\rightarrow j}$ in the same topology as in (a) as a function
    of shortest path distance $d(s,j)$.
    The decay has a wide spread due to direction dependence as explained in \eqref{eq:dipole-pot}.
    (c) Color coded plot of $\chi_{(st) \rightarrow j}$ in the
    Continental European Transmission Network topology with
    uniform free capacities (taken from \cite{Hutc13}). The vertex
    positions are not realistic.
    One central edge was 
    was perturbed. There are several highly susceptible vertices in the
    network periphery (dark blue). (d) Decay behavior of 
    $\chi_{(st)\rightarrow j}$ in the same topology as in (c) as a function
    of $d(s,j)$. The same edge as in (c) was perturbed. The susceptibilities 
    for a single distance are even more widely distributed than in a regular lattice 
    \label{fig:ev-susceptibility}.
    The straight lines in (b) and (d) are algebraic fits to the upper and lower envelopes of the data set to obtain the exponent of the power law decay. As the power law decay breaks down near the boundary due to finite size effects we have to choose a cutoff, restricting the fit to the shaded region.
    }
\end{figure}

Analogously to the case of vertex perturbation, the effects of edge 
perturbations can also be solved in the continuum limit, the result being 
the same as the potential due to an electrical dipole:
\begin{align}
    \xi(\vec{x}) &\propto \frac{\vec q \cdot \vec x}{|\vec x - \vec x_0|^2},
    \label{eq:dipole-pot}
\end{align}
where $\vec{q}$ is the unit vector in the direction along which the perturbed edge 
lie. 

This equation shows that unlike the response to vertex perturbation, 
the response to edge perturbation in a network will be highly \emph{directional}. 
The susceptibilities should decay the fastest in the direction \emph{along the edge perturbed}, 
according to the power law $d^{-2}$, consistent with the results presented in 
\cite{jungpre2016}, but much slower in the orthogonal direction.
In Fig.~\ref{fig:ev-susceptibility} (a,c), we see this direction dependence in
a regular square lattice. The lower envelope of the distance-susceptibility plot 
decays approximately as $d^{-2}$ as expected.

We repeat the same analysis on the Continental European Transmission Network. We see 
that the susceptibilities are spread even wider for constant distance, indicating a stronger 
dependence on the orientation of the edge. 
We notice that the upper envelope in Fig.~\ref{fig:ev-susceptibility} (d) decays very 
slowly: $\approx d^{-0.4}$, i.e.
there exists a small but nonzero number of nodes that are heavily affected by the perturbation, 
despite being very far away from the perturbed edge.

\subsection{Explaining the vulnerability of dead ends}
\label{sec-deadend}

The topology of a supply networks determines its local 
\cite{Peco98,Mott13} as well as global stability 
\cite{Menc13,12powergrid,12braess}.
Recently, Menck et al.~have shown that dead ends are particularly 
prone to instabilities \cite{Menc14}. They have measured  
the robustness of a power grid model to large perturbations 
at a single node in terms of the so-called basin stability. To this
end, the dynamics is simulated after a random perturbation to 
the steady state at a single node of the network.
The basin stability is then  defined as the probability 
that the network relaxes back to the steady state. Extensive 
Monte Carlo studies show that nodes adjacent to a dead
end or dead tree have a particularly small basin stability.

\begin{figure}[tb]
\centering
\includegraphics[height=4.4cm]{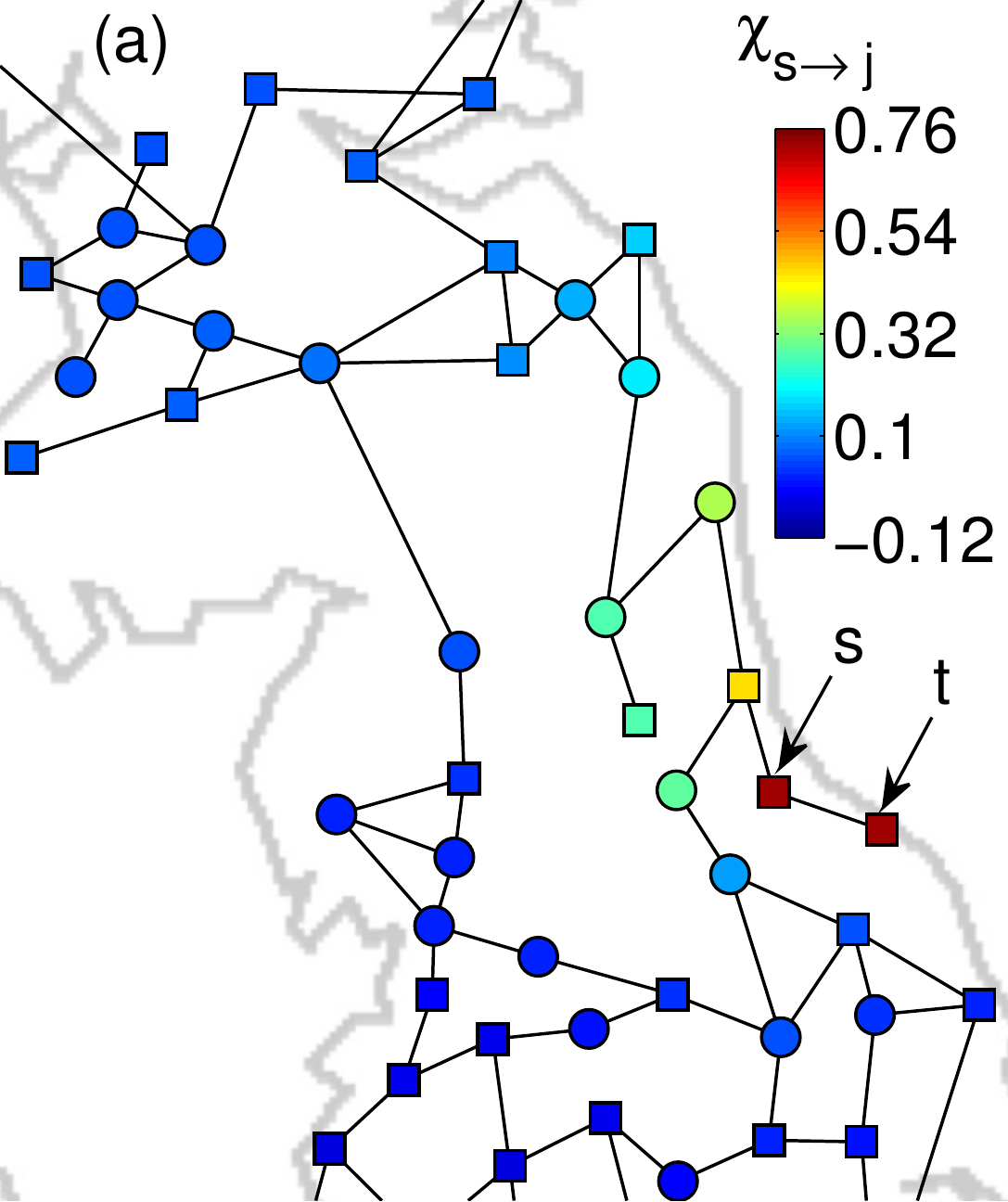}
\hspace{2mm}
\includegraphics[height=4.4cm]{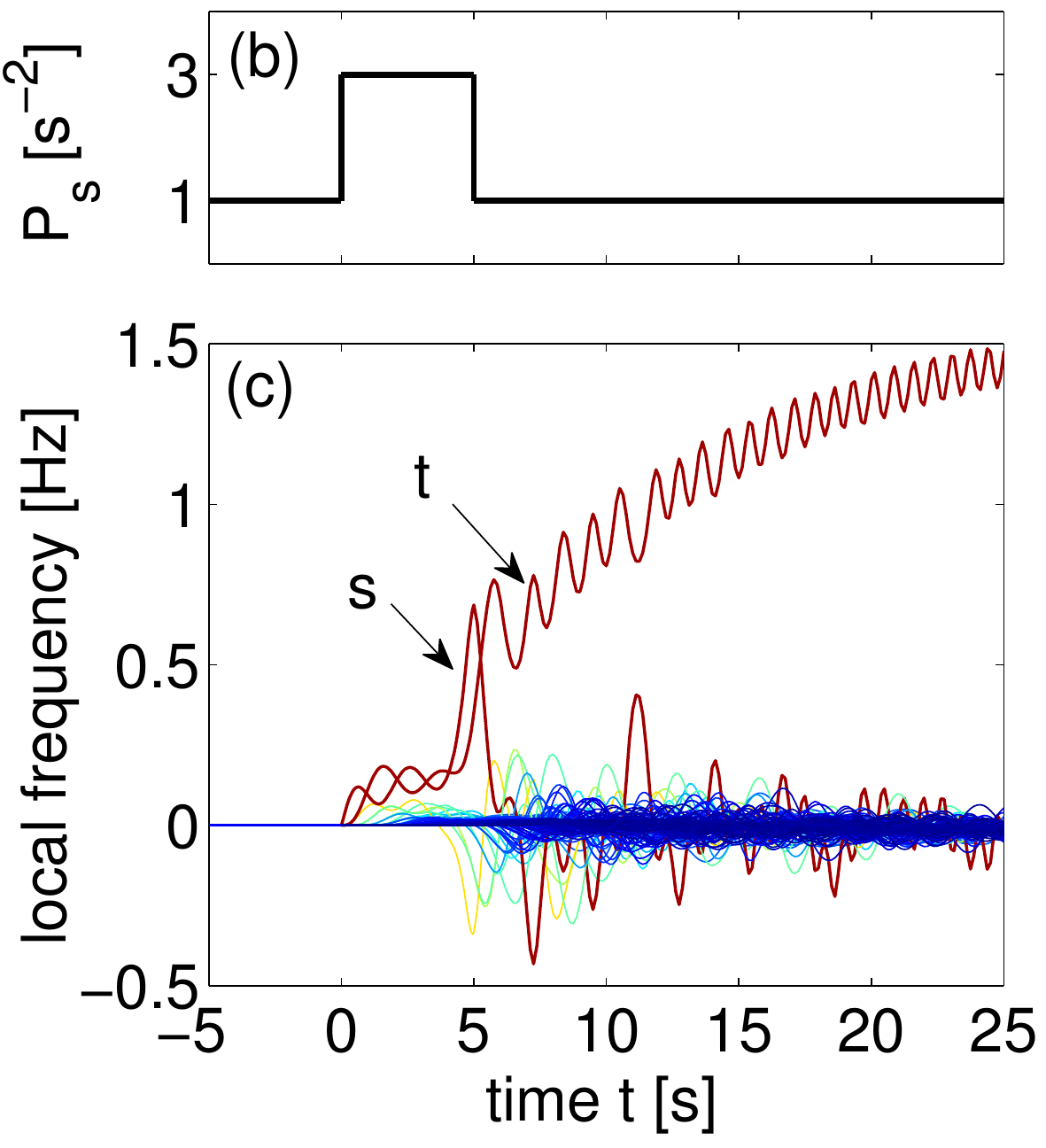}
\caption{
\label{fig:deadend}
Large susceptibility and desynchronization at dead ends.
(a) The edge susceptibility $\chi_{s \rightarrow j}$ is large
for $j=s$ and $j=t$, where $t$ is the dead end adjacent to the 
vertex $s$.
(b,c) Dynamics after a transient increase of the power 
$P_s$ at the vertex $s$. The impact of the perturbation
is strongest for the vertices $s$ and $t$. While $s$ 
relaxes after a short transient period, the dead end 
$t$ looses synchrony permanently.
We consider the topology of the British high-voltage transmission
grid as in Fig.~\ref{fig:globalsus}, of which only a
magnified part is shown. 
We assume that the moments of inertia $M_j\equiv M$ and the damping 
coefficient $D_j\equiv D$ are the same for all machines. The machines have
power injections of $P_j/M = \pm 1 \, {\rm s}^{-2}$ and all edges have the
transmission capacity $K/M = 4 \, {\rm s}^{-2}$ and $D/M = 0.1 \, {\rm s}^{-1}$.
% XXX
}
\end{figure}

The particular sensitivity of dead ends is directly related to the 
vertex-to-vertex susceptibility introduced in Sec.~\ref{sec:vertex-sus}.
The main mechanism causing desynchronization at
a dead end is shown in Fig.~\ref{fig:deadend}. The generation or power 
injection $P_s$ 
at a vertex $s$ adjacent to a dead end is increased for a short period 
of time. This perturbation has a strong influence on the vertex $s$ itself 
but also at the dead end $t$, causing a transient loss of  synchrony.
For longer times, the vertex $s$ relaxes and resynchronizes 
with the rest of the network, whereas the dead end $t$ does not. 
In summary, a perturbation at the vertex $s$ has a large influence on
the dynamics of the dead end, while its influence on the 
bulk of the network is small.

This property if directly mirrored by the vertex-to-vertex
susceptibility $\chi_{s \rightarrow j}$. Generally, the susceptibility
is largest locally, i.e.~ for $j=s$, while it decreases 
with the distance as discussed above. Only if $s$ is adjacent to a dead end $t$, the 
non-local susceptibility $\chi_{s \rightarrow t}$ is comparably
large. One particular example is shown in Fig.~\ref{fig:deadend} (a).
This shows that the non-local impact is strongest at dead ends
and thus provides an explanation for their low basin stability.
\begin{figure*}[!htb]
\centering
\includegraphics[width=14cm]{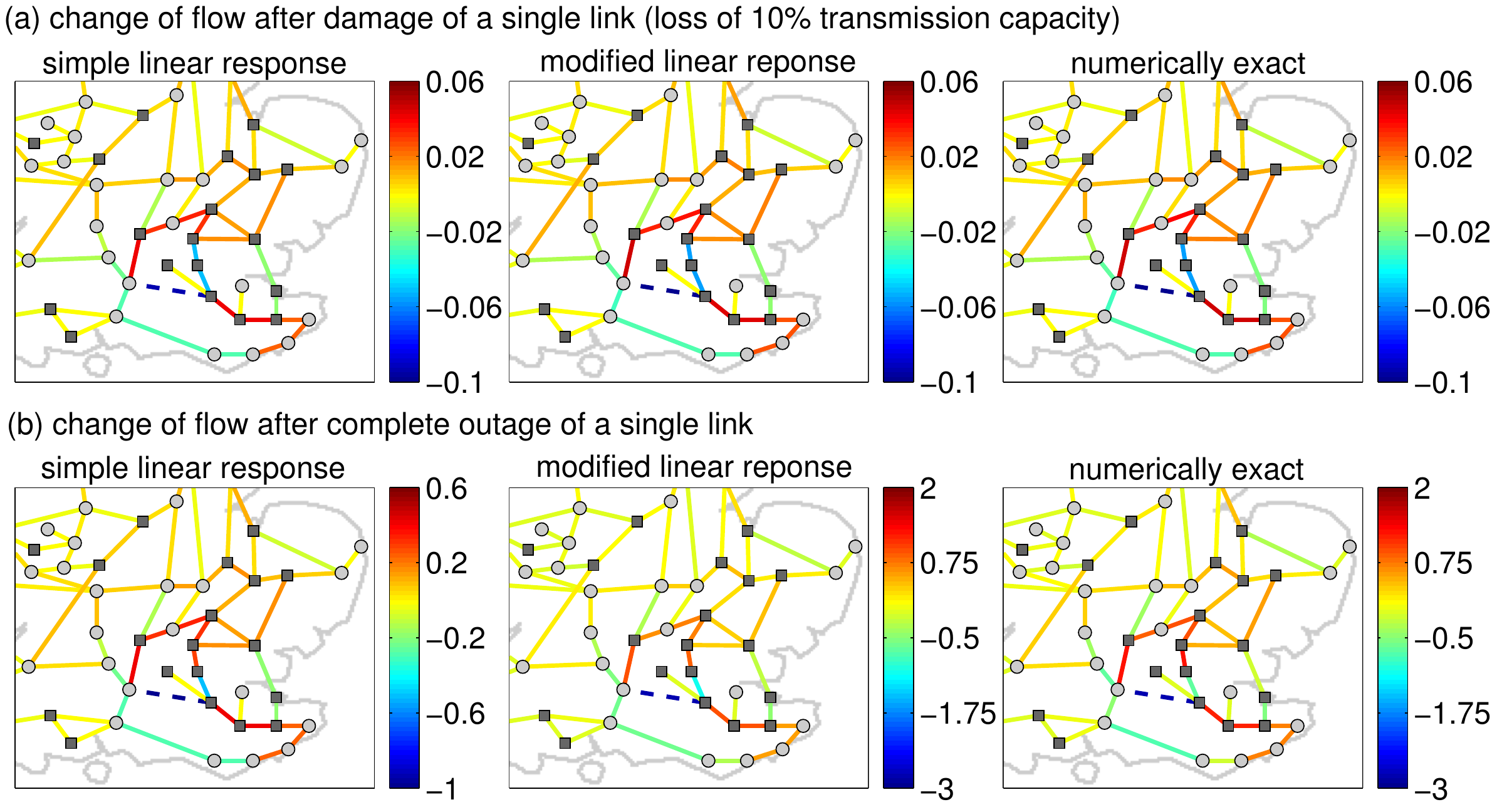}
\caption{
\label{fig:flowchange}
Change of flow magnitudes $|F_{ij}|$ after (a) the damage ($\kappa = -0.1 \times K_{st}$) or 
(b) the complete outage ($\kappa = - K_{st}$) of a single edge (dashed).
We compare the prediction of simple linear response approach (\ref{eqn:flow-edge}) and the modified formula (\ref{eqn:flow-edge-re}) to the results of a numerical solution of the steady state condition (\ref{eqn:def-steady}). Note the different color scales used in the figure.
We consider the topology British high-voltage transmission grid as in Fig.~\ref{fig:globalsus}, of which only a
magnified part is shown. 
% Simulation and plotting: critical_predict2.m
}
\end{figure*}

\section{Large perturbations and structural damages}
\label{sec:crit-lines}
\subsection{From small to large changes}
\label{sec:line-out}
Linear response theory readily predicts how the flow in a network changes after a small perturbation of the network topology. But can it be used to estimate the effects of major changes such as the complete outage of an edge? This is especially important for electric power grids, where transmission line failures repeatedly induce large-scale outages (see, e.g., \cite{Albe00,Cohe01,Mott02,Albe04,Buld10,Fair04,Pour06}).
Thus any method that helps to predict the stability of a grid after the failure of a single edge is extremely valuable. For an ad-hoc analysis of network stability in practical applications such methods should be only based on the topological and load properties of the original network and avoid time consuming direct numerical simulations.

We can treat macroscopic changes within a linear response approach if we slightly modify the derivation of edge susceptibilities introduced in Sec.~\ref{sec:pert-edge}. As before we keep only terms linear in $\vec \xi$ but we drop the assumption that the perturbations $\kappa_{ij}$ are small. Then Eq.~(\ref{eqn:steady2}) has to be modified as 
\bea
   && \sum_{i=1}^N (K_{ij}+\kappa_{ij}) \cos(\phi_i-\phi_j) (\xi_i-\xi_j) \nn \\
         && \qquad  = -     
      \sum_{i=1}^N \kappa_{ij} \sin(\phi_i-\phi_j).    \nn \\
\eea
This set of linear equation is rewritten in matrix form as
\be
   A_{(st)} \vec \xi = \kappa L_{st} \vec q_{(st)}
   \label{eq:steady4}
\ee
with the matrix
\be
   A_{(st)} = A + \kappa \cos(\phi_s - \phi_t) \, \vec q_{(st)}  \vec q_{(st)}^T,
\ee
where the superscript $T$ denotes the transpose of a vector or matrix. The change of the local phases is then obtained by formally solving Eq.~(\ref{eq:steady4}),
\be
    \vec \xi = \kappa L_{st} A_{(st)}^+ \vec q_{(st)}.
\ee
In particular we will need the phase differences between two nodes which is given as
\be
   \xi_j - \xi_i = \kappa L_{st}  \vec q_{(ji)}^T  A_{(st)}^+ \vec q_{(st)}.
\ee
This expression suggests that we need to calculate the inverse separately for each edge $(s,t)$ if we want to assess the impact of all possible edge failures. However, we can greatly simplify the problem using the Woodbury matrix identity \cite{Wood50}, which yields
\begin{align*}
   A_{(st)}^+ &=  (A + \kappa \cos(\phi_s - \phi_t) \,  \vec q_{(st)}  \vec q_{(st)}^T)^+  \nn \\
   & = A^+ - A^+ \vec q_{(st)} (\kappa^{-1} + \vec q_{(st)}^T A^+ \vec q_{(st)})^+ \vec q_{(st)}^T A^+
\end{align*}
We then obtain
%\begin{align*}
%    & \vec q_{(ji)}^T  A_{(st)}^+  \vec q_{(st)} \\
%    &= \vec q_{(ji)}^T  A^+  \vec q_{(st)}   \\
%    & \quad - \vec q_{(ji)}^T   A^+ \vec q_{(st)} (\kappa^{-1} \cos(\phi_s - \phi_t)^{-1} \, + \vec q_{(st)}^T A^+ \vec q_{(st)})^+ \vec q_{(st)}^T A^+  \vec q_{(st)} \\
%    & = \frac{\vec q_{(ji)}^T  A^+  \vec q_{(st)} }{1+ \kappa \cos(\phi_s - \phi_t) \, \vec q_{(st)}^T  A^+  \vec q_{(st)} } \, .
%\end{align*}
\be
    \vec q_{(ji)}^T  A_{(st)}^+  \vec q_{(st)} 
     = \frac{\vec q_{(ji)}^T  A^+  \vec q_{(st)} }{1+ \kappa \cos(\phi_s - \phi_t) \, \vec q_{(st)}^T  A^+  \vec q_{(st)} } \, .
\ee

The network flows after the perturbation are now given by
\begin{align}
    F''_{ij} & = K_{ij}\sin{(\varphi_j - \varphi_i + \xi_j - \xi_i)} \nn \\
             & = F_{ij} + K_{ij}\cos{(\varphi_j - \varphi_i)}(\xi_j - 
    \xi_i)\nn \\
             & = F_{ij} + \widetilde K_{ij}\kappa L_{st}  \vec q_{(ji)}^T  
    A_{(st)}^+ \vec q_{(st)} \nn \\
             & = F_{ij} +    
          \, \frac{\kappa L_{st} \widetilde K_{ij} (T_{js} - T_{jt} - T_{is} + T_{it})}{
               1 + \kappa \cos(\phi_s - \phi_t) \, (T_{ss} - T_{st} - T_{ts} + T_{tt})} \nn
\end{align}
for all edges $(i,j) \neq (s,t)$. This expression differs from Eq.~(\ref{eqn:flow-edge}) only by the denominator which tends to one in the limit of small perturbations $\kappa \rightarrow 0$. 
For a macroscopic perturbation the denominator is essential to predict the magnitude of the flow changes correctly. The complete failure of an edge is described by $\kappa = -K_{st}$ such that we obtain
\begin{align}
    F''_{ij} & = F_{ij} - 
          \, \frac{\widetilde K_{ij}    (T_{bs} - T_{bt} - T_{as} + T_{at})}{
               1 - \widetilde K_{st} (T_{ss} - T_{st} - T_{ts} + T_{tt})} \times F_{st} 
               \label{eqn:flow-edge-re}
\end{align}
for all  edges $(i,j) \neq (s,t)$ and $F''_{st} = 0$ for the failed edge. 
Similar formulae are used in power engineering, where the fraction is referred 
to as a Line Outage Distribution Factor (LODF)  \cite{Wood13, LODF2016}.

An example of how the damage of a single transmission line affects the flows in a power grid is shown in Fig.~\ref{fig:flowchange}. We plot the change of the flow magnitude $|F_{ij}|$ predicted by the simple linear response approach (\ref{eqn:flow-edge}) and the modified approach (\ref{eqn:flow-edge-re}) in comparison to the actual value obtained from a numerical solution of the
steady state condition (\ref{eqn:def-steady}).
For a small damage where only $10 \%$ of the transmission capacity is lost ($\kappa_{st} = - 0.1 \times K_{st}$) we find a very good agreement between the predicted and actual values as expected. But even in for a complete breakdown the modified formula (\ref{eqn:flow-edge-re}) provides a very good prediction of the flow changes after the damage. The simpler linear response formula (\ref{eqn:flow-edge}) strongly underestimates the flow changes as it neglects the denominator $1 - \widetilde K_{st} (T_{ss} - T_{st} - T_{ts} + T_{tt})$, which is significantly smaller than one in the current example.

\begin{figure*}[tb]
\centering
\includegraphics[width=12.9cm]{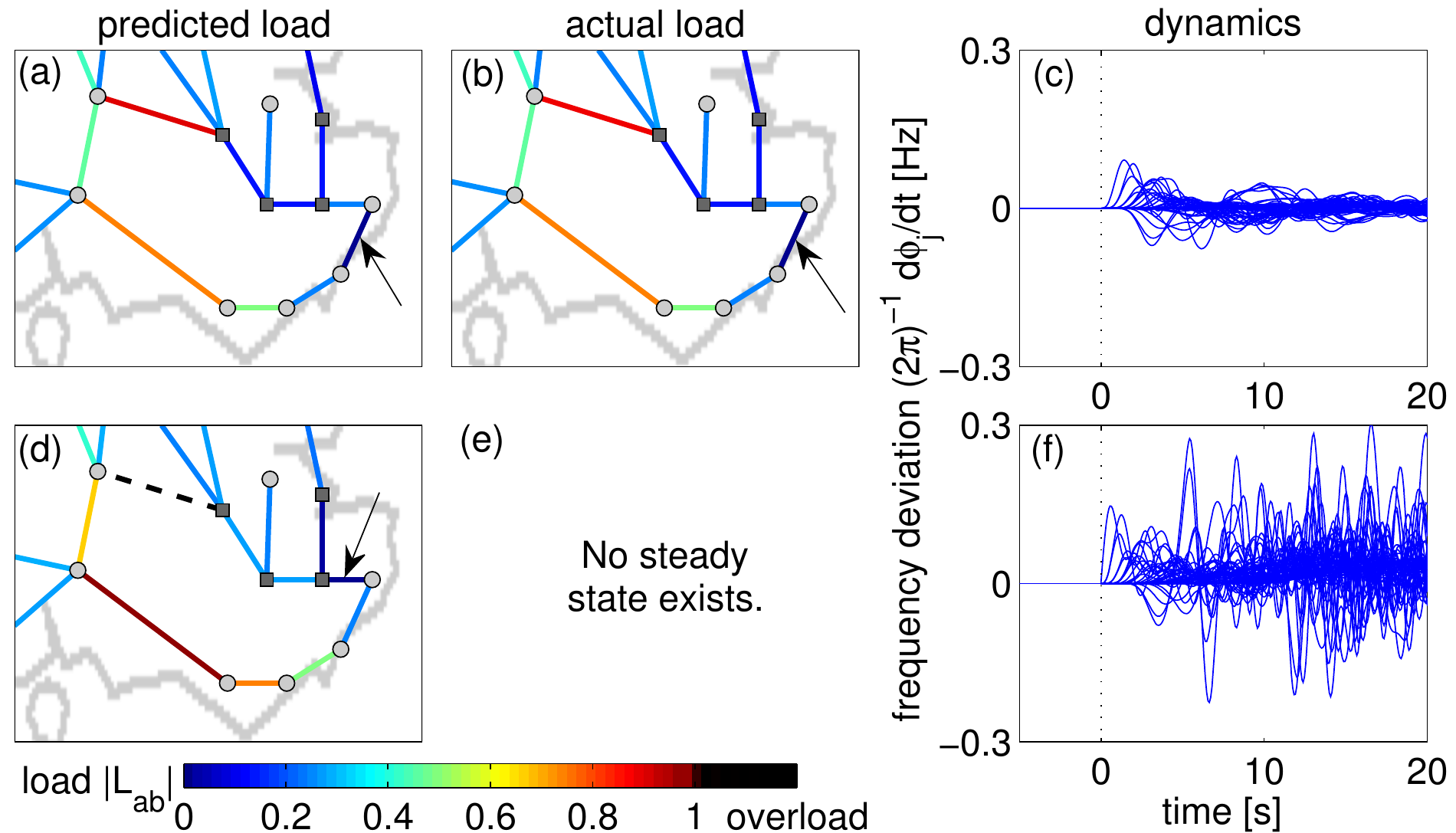}
\caption{
\label{fig:crtilinks}
Identification of critical edges using linear response theory.
We analyze the effects of the breakdown of a single transmission line
for two examples marked by arrows. 
(a-c) In the upper example, no secondary overloads occur and the 
grid relaxes back to steady operation after a short transient period.
(d-f) In the lower example, linear
response theory predicts a secondary overload (black dashed line in d)
and consequently the dynamics becomes unstable as shown in panel (f).
(a,d) Loads $|F_{ij}''/K_{ij}|$ predicted
by the modified linear response formula  (\ref{eqn:flow-edge-re})
(b,e) Actual load obtained by solving the steady state 
condition (\ref{eqn:def-steady}). In (e) no steady state exists
after the initial breakdown.
(c,f) Grid dynamic obtained after the breakdown of the respective edge
at $t=0$.
We consider the topology of the British high-voltage transmission
grid as in Fig.~\ref{fig:globalsus}, of which only a
magnified part is shown. 
We assume that the moments of inertia $M_j\equiv M$ and the damping 
coefficient $D_j\equiv D$ are the same for all machines. The machines have
power injections of $P_j/M = \pm 1 \, {\rm s}^{-2}$ and all edges have the
transmission capacity $K/M = 4 \, {\rm s}^{-2}$ and $D/M = 0.1 \, {\rm s}^{-1}$.
% Simulation and plotting: critical_loaddyn1.m
}
\end{figure*}

\subsection{Identification of critical edges}

The modified formula (\ref{eqn:flow-edge-re}) can be used to predict impeding
overloads and large scale outages in complex supply networks \cite{16critical}. 
Figure \ref{fig:crtilinks} shows the effect of the  breakdown of a single transmission 
line for two examples. In the first example, formula  (\ref{eqn:flow-edge-re}) predicts that
no overload occurs in agreement with the direct solution of the steady state 
condition (\ref{eqn:def-steady}). Thus we
expect that the grid relaxes to a new steady state after the
failure of the respective edge. This prediction is confirmed by
a direct numerical simulation of the equations of motion
(\ref{eqn:eom-theta}). 
In the second example  formula  (\ref{eqn:flow-edge-re}) predicts that
further overloads occur, i.e. that $|F''_{ij}/K_{ij}| > 1$ for at least
one edge $(i,j)$, after the transmission line $(s,t)$ failed. Indeed, 
numerical simulations show that no steady state solution of  
Eq.~(\ref{eqn:def-steady}) exists and that the grid becomes unstable and 
looses synchrony. 
In the following, we call an edge ``critical'' if its breakdown 
induces a desynchronization of the grid. If the grid relaxes back to a 
steady operation, i.e. an attractively stable synchronized state 
with $\dot \phi_j = 0$ for all $j$, we call the edge ``stable''.

\begin{figure}[tb]
\centering
\includegraphics[width=8cm]{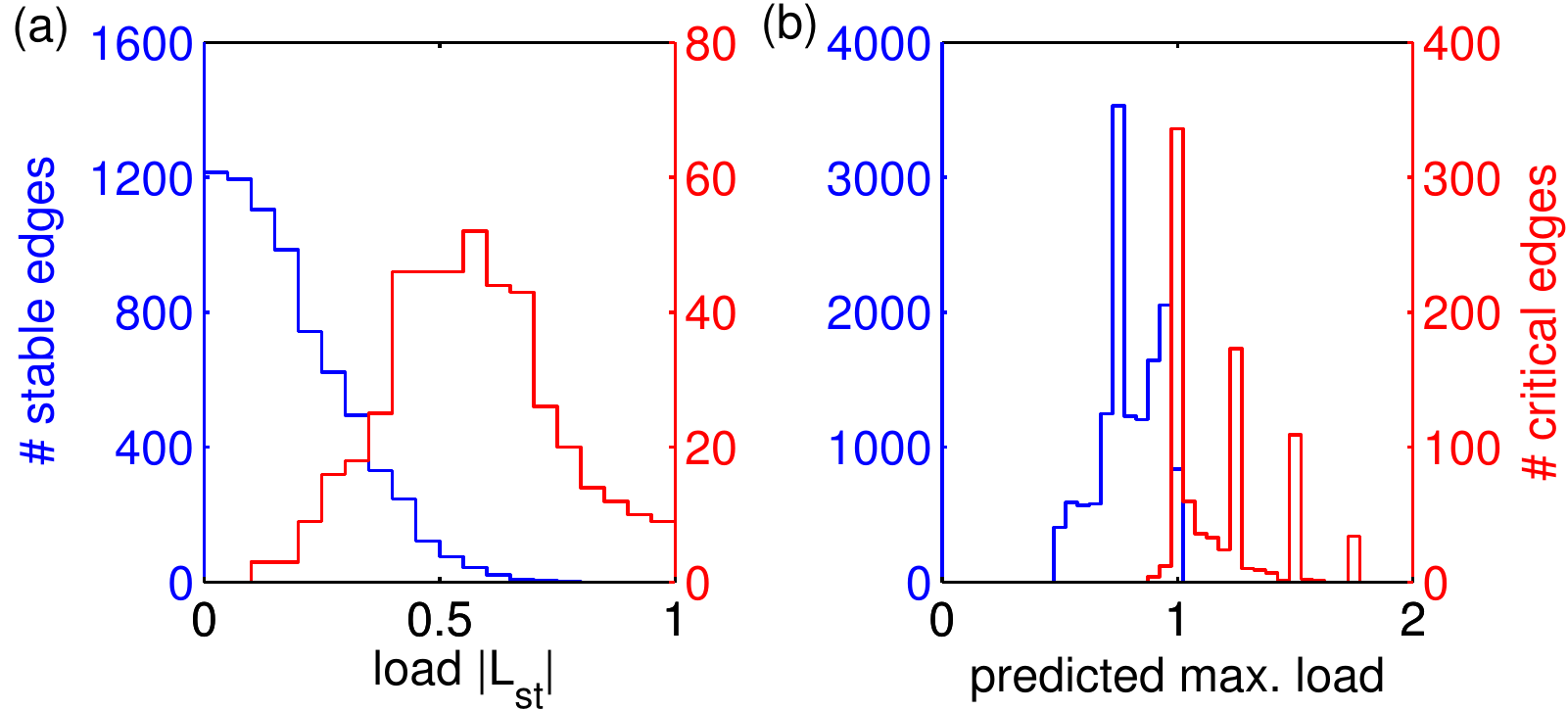}
\includegraphics[width=8cm]{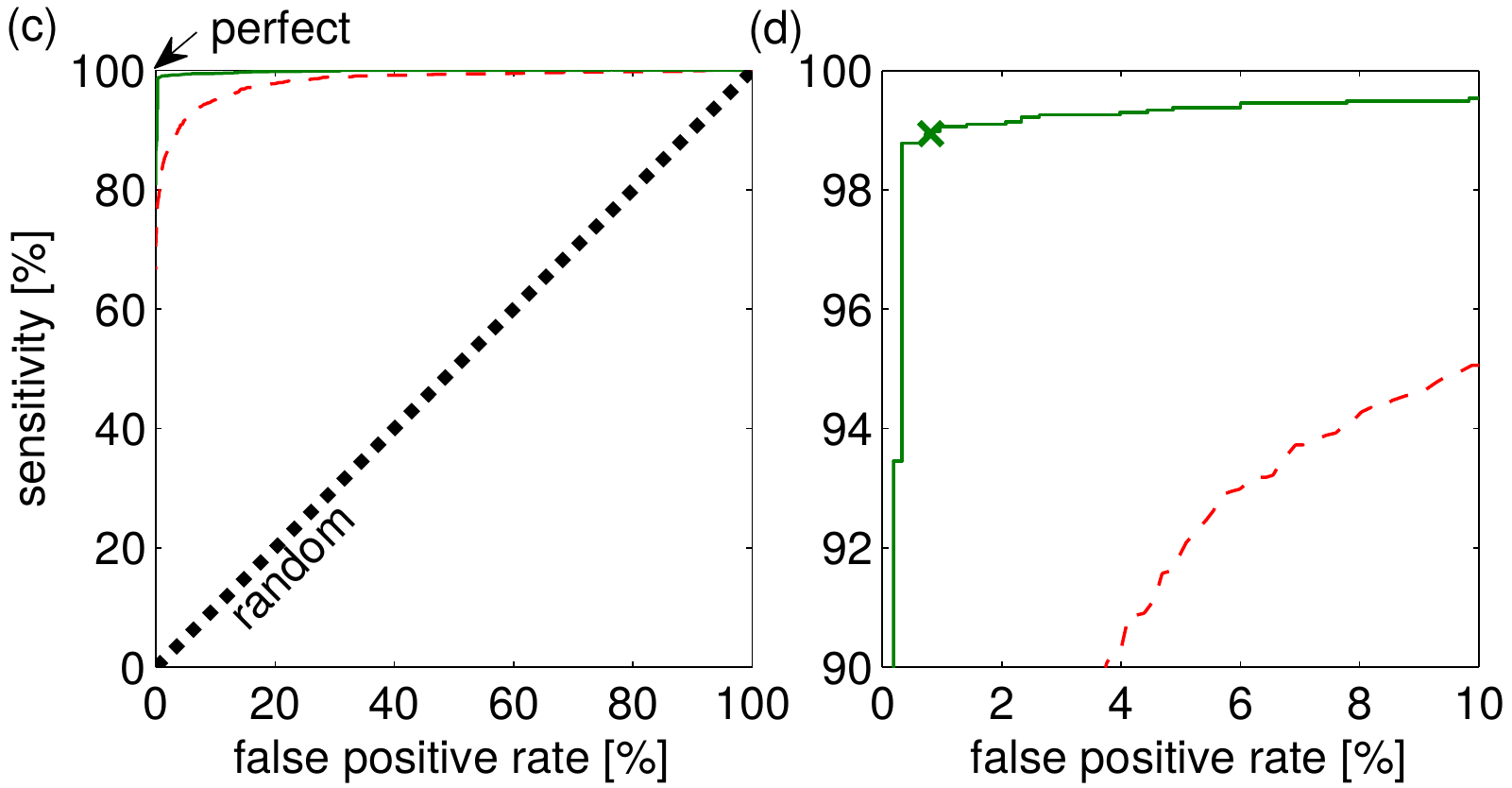}
\caption{
\label{fig:histroc}
Performance of different classifiers for the prediction of
critical edges.
(a,b) Histograms of characteristic quantities to identify critical 
(red) and stable (blue) edges in complex supply networks:
(a) load $|L_{st}|$ before breakdown and
(b) the maximum load $\max_{(i,j)} |F''_{ij}/K_{ij}|$ predicted by
linear response theory.  
(c,d) The performance of the classifiers can be judged by a receiver operating 
characteristics (ROC) curve, where the sensitivity is plotted vs. the false positive 
rate for different threshold values $h$. The predicted max.~load (solid green line) 
closely approaches the perfect limit $(0,1)$ and clearly outperforms a classifier 
based on the load $|L_{st}|$ (dashed).
Results are collected for 100 realizations of the British grid with
random positions of generators and consumers. One realization
is shown in Fig.~\ref{fig:flowchange} and Fig.~\ref{fig:crtilinks}.
}
\end{figure}

Based on these results we propose to use the maximum load
$\max_{(i,j)} |F''_{ij}/K_{ij}|$ predicted by the modified linear
response formula  (\ref{eqn:flow-edge-re}) as a criterion to infer 
network stability. An edge $(s,t)$ is predicted to be ``critical''
or ``stable'' according to the following classification system:
\bea
   && \max_{(i,j)} |F''_{ij}/K_{ij}| >    h   \;  \Rightarrow \;  \mbox{predicted to be critical}, \nn \\
   && \max_{(i,j)} |F''_{ij}/K_{ij}| \le h   \; \Rightarrow \;  \mbox{predicted to be stable},
   \label{eqn:def-class2}
\eea
where $h$ is a threshold value. Bridges, i.e. edges whose removal disconnects 
the grid are always predicted to be critical.

To test this method, we perform direct numerical simulations of the equations 
of motion (\ref{eqn:eom-theta}) for a large number of test grids, each starting from a
stationary state of normal operation and study the  influence of the breakdown of a 
single edge.
Examples for both scenarios are shown in Fig.~\ref{fig:crtilinks}.
We analyze the coarse-grained  structure of the British high-voltage 
transmission grid \cite{Simo08,12powergrid} which has 165 edges.
We consider 100 random realizations with random generator 
positions, thus testing  $16\,500$ edges in total. 
For each out of 100 random realizations, we fix the network
topology by randomly selecting half of the nodes to be generators 
($P_j = + 1 \, P_0$) and the others to be consumers 
($P_j = - P_0$), with $P_0 = 1\, {\rm s}^{-2}$. The transmission
capacity of all edges is fixed as $K_{ij} = K_0 =4 \, {\rm s}^{-2}$.
One example of such a network is depicted in Fig.~\ref{fig:globalsus}.
Networks not supporting a steady state before any edge 
breakdown were discarded.

To evaluate the performance of the proposed classification scheme 
(\ref{eqn:def-class2}) we must first
define the possible outcomes of a prediction, where we distinguish
between two different kinds of prediction errors:\\
\phantom{0} \emph{True positive:} Edge is predicted critical and is critical.\\
\phantom{0} \emph{False positive:} Edge is predicted critical but is stable.\\
\phantom{0} \emph{False negative:} Edge is predicted stable but is critical.\\
\phantom{0} \emph{True negative:} Edge is predicted stable and is stable.\\
Generally it is impossible to rule out both false negative and false
positive predictions such that a compromise must be achieved.
In the current setting, the number of false positive predictions 
can be minimized by choosing a high value of $h$, while the number
of false negative predictions can be minimized by choosing 
a small value of $h$.

A quantitative assessment of the performance of a classifiers is then provided 
by a receiver operating characteristics (ROC) curve (Fig.~\ref{fig:histroc}) \cite{Egan75}. 
Here, the true positive rate of the test, also called the sensitivity
\be
    \mbox{SEN} := \frac{ 
    \#\mbox{true positive predictions}
    }{ 
    \#\mbox{true positive pred.} +  \#\mbox{false negative pred.}
   }  \nn
  \label{eqn:def-sen}
\ee
is plotted vs. the false positive rate 
\be
    \mbox{FPR} := \frac{ 
    \#\mbox{false positive predictions}
    }{ 
    \#\mbox{false positive pred.} +  \#\mbox{true negative pred.}
   }  \nn
\ee
for different threshold values $h$. For a perfect classifier, the ROC 
is a point at $(\mbox{FPR},\mbox{SEN}) =  (0,1)$, while for a fully random 
classification the ROC curve is a straight line with slope $1$ through the 
origin. Therefore, a classifier is judged to be the better, the nearer 
the ROC curve approaches the point $(0,1)$, i.e. the upper left corner 
of the plot. 

Numerical results for 100 realization of the British grid with random
generator positions are shown in Fig.~\ref{fig:histroc} (c,d). It is
observed that the the classifier (\ref{eqn:def-class2}) closely approaches the 
perfect limit  $(\mbox{FPR},\mbox{SEN}) =  (0,1)$ and clearly
outperforms a classifier based on the load of the edges.
Therefore linear response theory provides a very promising
approach to identifying critical infrastructures in complex supply
networks.

\section{Conclusion and outlook}
In summary, we introduced the concepts of \textit{vertex susceptibilities} and 
\textit{edge susceptibilities} as measures of responses to parametric changes 
in network dynamical systems.   They qualitatively distinguish -- and quantify 
-- the responses due changes in the properties of units and their 
interactions, respectively.   Focusing on steady state responses of oscillator 
network characterized by phases or phases and their velocities, we derived 
explicit forms of such \textit{network susceptibilities}. We in particular 
analyzed the role of irregular interaction topologies as those are the least 
investigated compared to the susceptibilities that are standard in physics.  
Specifically,  we have analyzed how the responses of a network in some given phase
locked state depend on the relative location of perturbation and response 
sites and how the network topologies enter.   We linked susceptibilities to 
established measures, for instance, special cases are known as line outage 
distribution factors in power grid engineering and susceptibilities are 
closely related to centrality measures.   We explicated an accurate prediction 
of network responses not only to small perturbation but also after the full 
breakdown of edges.   In power grids, this may be applied, for instance, for 
an ad-hoc security assessment.   Furthermore, network susceptibilities
directly reveal weak points of flow networks and may thus be used in the 
planning and design of future grid extensions and establishing other supply 
network infrastructures.   Finally, the two types of network susceptibilities 
are generic measures of responses to parameter changes and as such may be 
straightforwardly generalized across flow, transport and supply networks as 
well as other network dynamical systems where responses are nonlocal due to 
genuine collective dynamics.

\acknowledgments

We gratefully acknowledge support from the Federal Ministry of Education and Research (BMBF grant no.~03SF0472A-E),  the Deutsche Forschungsgemeinschaft (DFG grant no. ME-1332/19-1), the Helmholtz Association (via the joint initiative ``Energy System 2050 -- A Contribution of the Research Field Energy'' and the grant no.~VH-NG-1025 to D.W.) and the Max Planck Society to M.T.
The works of D.M., H.R. and X.Z. are supported by the IMPRS Physics of Biological and Complex Systems, G\"ottingen.

\begin{appendix}
\section{Global edge-to-vertex susceptibility in terms of resistance distance}
\label{app:global-susc}

We start with Eq.~(\ref{eq:R_T_relation})
\begin{align}
    T_{ij} & = -\frac{1}{2}R_{ij} +\frac{1}{2N}\left(R_j^{\rm tot} +R_i^{\rm tot}\right) -\frac{\sum_{i,j}R_{ij}}{N^2}.
\end{align}
Applying this on Eq.~(\ref{eq:global-edge-susc}), we obtain
\begin{align*}
    \frac{\chi^2_{(st)}}{L_{st}^2} & =  \sum_{u=1}^N  ( T_{us} - T_{ut})^2 \\
                  & = \sum_{u=1}^N \left[ -\frac{1}{2}(R_{us}-R_{ut}) 
+ \frac{1}{2N}(R_{us}^{\rm tot}-R_{ut}^{\rm tot}) \right]^2\\
& =  \frac{1}{4N}\left[(R_s^{\rm tot}-R_t^{\rm tot})\right]^2
+ \frac{1}{4}\sum_u (R_{us}-R_{ut})^2\\
& \qquad - \frac{1}{2N}(R_s^{\rm tot}-R_t^{\rm tot})\sum_{u}(R_{us} - R_{ut}) \\
& =  \frac{1}{4N}\left[(R_s^{\rm tot}-R_t^{\rm tot})\right]^2
+ \frac{1}{4}\sum_u (R_{us}-R_{ut})^2\\ 
& \qquad - \frac{1}{2N}\left[(R_s^{\rm tot}-R_t^{\rm tot})\right]^2 \\
& =  -\frac{1}{4N}(R_s^{\rm tot}-R_t^{\rm tot})^2+\frac{1}{4}\sum_u(R_{us}-R_{ut})^2\\
& =  \frac{N}{4} \left[ 
\frac{1}{N}\sum_u(R_{us}-R_{ut})^2-\left[\frac{1}{N}\sum_uR_{us - R_{ut}}\right]^2\right].\\
\end{align*}

\end{appendix}

% --- Literatur -------------------------------------------------------------------

\bibliography{centrality}

\begin{thebibliography}{44}
\expandafter\ifx\csname natexlab\endcsname\relax\def\natexlab#1{#1}\fi
\expandafter\ifx\csname bibnamefont\endcsname\relax
  \def\bibnamefont#1{#1}\fi
\expandafter\ifx\csname bibfnamefont\endcsname\relax
  \def\bibfnamefont#1{#1}\fi
\expandafter\ifx\csname citenamefont\endcsname\relax
  \def\citenamefont#1{#1}\fi
\expandafter\ifx\csname url\endcsname\relax
  \def\url#1{\texttt{#1}}\fi
\expandafter\ifx\csname urlprefix\endcsname\relax\def\urlprefix{URL }\fi
\providecommand{\bibinfo}[2]{#2}
\providecommand{\eprint}[2][]{\url{#2}}

\bibitem[{\citenamefont{Strogatz}(2001)}]{Stro01}
\bibinfo{author}{\bibfnamefont{S.~H.} \bibnamefont{Strogatz}},
  \bibinfo{journal}{Nature} \textbf{\bibinfo{volume}{410}},
  \bibinfo{pages}{268} (\bibinfo{year}{2001}).

\bibitem[{\citenamefont{Newman}(2010)}]{Newm10}
\bibinfo{author}{\bibfnamefont{M.~E.~J.} \bibnamefont{Newman}},
  \emph{\bibinfo{title}{Networks -- An Introduction}}
  (\bibinfo{publisher}{Oxford University Press}, \bibinfo{address}{Oxford},
  \bibinfo{year}{2010}).

\bibitem[{\citenamefont{Kuramoto}(1975)}]{Kura75}
\bibinfo{author}{\bibfnamefont{Y.}~\bibnamefont{Kuramoto}}, in
  \emph{\bibinfo{booktitle}{International Symposium on on Mathematical Problems
  in Theoretical Physics}}, edited by
  \bibinfo{editor}{\bibfnamefont{H.}~\bibnamefont{Araki}}
  (\bibinfo{publisher}{Springer}, \bibinfo{address}{New York},
  \bibinfo{year}{1975}), Lecture Notes in Physics Vol. 39, p.
  \bibinfo{pages}{420}.

\bibitem[{\citenamefont{Kuramoto}(1984)}]{Kura84}
\bibinfo{author}{\bibfnamefont{Y.}~\bibnamefont{Kuramoto}},
  \emph{\bibinfo{title}{Chemical Oscillations, Waves, and Turbulence}}
  (\bibinfo{publisher}{Springer}, \bibinfo{address}{Berlin},
  \bibinfo{year}{1984}).

\bibitem[{\citenamefont{Sompolinsky et~al.}(1990)\citenamefont{Sompolinsky,
  Golomb, and Kleinfeld}}]{Somp90}
\bibinfo{author}{\bibfnamefont{H.}~\bibnamefont{Sompolinsky}},
  \bibinfo{author}{\bibfnamefont{D.}~\bibnamefont{Golomb}}, \bibnamefont{and}
  \bibinfo{author}{\bibfnamefont{D.}~\bibnamefont{Kleinfeld}},
  \bibinfo{journal}{Proc. Natl. Acad. Sci. U.S.A.}
  \textbf{\bibinfo{volume}{87}}, \bibinfo{pages}{7200} (\bibinfo{year}{1990}).

\bibitem[{\citenamefont{Wiesenfeld et~al.}(1996)\citenamefont{Wiesenfeld,
  Colet, and Strogatz}}]{Wies96}
\bibinfo{author}{\bibfnamefont{K.}~\bibnamefont{Wiesenfeld}},
  \bibinfo{author}{\bibfnamefont{P.}~\bibnamefont{Colet}}, \bibnamefont{and}
  \bibinfo{author}{\bibfnamefont{S.~H.} \bibnamefont{Strogatz}},
  \bibinfo{journal}{Phys. Rev. Lett.} \textbf{\bibinfo{volume}{76}},
  \bibinfo{pages}{404} (\bibinfo{year}{1996}).

\bibitem[{\citenamefont{Vladimirov et~al.}(2003)\citenamefont{Vladimirov,
  Kozireff, and Mandel}}]{Vlad03}
\bibinfo{author}{\bibfnamefont{A.~G.} \bibnamefont{Vladimirov}},
  \bibinfo{author}{\bibfnamefont{G.}~\bibnamefont{Kozireff}}, \bibnamefont{and}
  \bibinfo{author}{\bibfnamefont{P.}~\bibnamefont{Mandel}},
  \bibinfo{journal}{Europhys. Lett.} \textbf{\bibinfo{volume}{61}},
  \bibinfo{pages}{613} (\bibinfo{year}{2003}).

\bibitem[{\citenamefont{Heinrich et~al.}(2011)\citenamefont{Heinrich, Ludwig,
  Qian, Kubala, and Marquardt}}]{Hein11}
\bibinfo{author}{\bibfnamefont{G.}~\bibnamefont{Heinrich}},
  \bibinfo{author}{\bibfnamefont{M.}~\bibnamefont{Ludwig}},
  \bibinfo{author}{\bibfnamefont{J.}~\bibnamefont{Qian}},
  \bibinfo{author}{\bibfnamefont{B.}~\bibnamefont{Kubala}}, \bibnamefont{and}
  \bibinfo{author}{\bibfnamefont{F.}~\bibnamefont{Marquardt}},
  \bibinfo{journal}{Phys. Rev. Lett.} \textbf{\bibinfo{volume}{107}},
  \bibinfo{pages}{043603} (\bibinfo{year}{2011}).

\bibitem[{\citenamefont{Rohden et~al.}(2012)\citenamefont{Rohden, Sorge, Timme,
  and Witthaut}}]{12powergrid}
\bibinfo{author}{\bibfnamefont{M.}~\bibnamefont{Rohden}},
  \bibinfo{author}{\bibfnamefont{A.}~\bibnamefont{Sorge}},
  \bibinfo{author}{\bibfnamefont{M.}~\bibnamefont{Timme}}, \bibnamefont{and}
  \bibinfo{author}{\bibfnamefont{D.}~\bibnamefont{Witthaut}},
  \bibinfo{journal}{Phys. Rev. Lett.} \textbf{\bibinfo{volume}{109}},
  \bibinfo{pages}{064101} (\bibinfo{year}{2012}).

\bibitem[{\citenamefont{Witthaut and Timme}(2012)}]{12braess}
\bibinfo{author}{\bibfnamefont{D.}~\bibnamefont{Witthaut}} \bibnamefont{and}
  \bibinfo{author}{\bibfnamefont{M.}~\bibnamefont{Timme}},
  \bibinfo{journal}{New J. Phys.} \textbf{\bibinfo{volume}{14}},
  \bibinfo{pages}{083036} (\bibinfo{year}{2012}).

\bibitem[{\citenamefont{D\"orfler et~al.}(2013)\citenamefont{D\"orfler,
  Chertkov, and Bullo}}]{Dorf13}
\bibinfo{author}{\bibfnamefont{F.}~\bibnamefont{D\"orfler}},
  \bibinfo{author}{\bibfnamefont{M.}~\bibnamefont{Chertkov}}, \bibnamefont{and}
  \bibinfo{author}{\bibfnamefont{F.}~\bibnamefont{Bullo}},
  \bibinfo{journal}{Proceedings of the National Academy of Sciences}
  \textbf{\bibinfo{volume}{110}}, \bibinfo{pages}{2005} (\bibinfo{year}{2013}).

\bibitem[{\citenamefont{Motter et~al.}(2013)\citenamefont{Motter, Myers,
  Anghel, and Nishikawa}}]{Mott13}
\bibinfo{author}{\bibfnamefont{A.~E.} \bibnamefont{Motter}},
  \bibinfo{author}{\bibfnamefont{S.~A.} \bibnamefont{Myers}},
  \bibinfo{author}{\bibfnamefont{M.}~\bibnamefont{Anghel}}, \bibnamefont{and}
  \bibinfo{author}{\bibfnamefont{T.}~\bibnamefont{Nishikawa}},
  \bibinfo{journal}{Nature Physics} \textbf{\bibinfo{volume}{9}},
  \bibinfo{pages}{191} (\bibinfo{year}{2013}).

\bibitem[{\citenamefont{Rohden et~al.}(2014)\citenamefont{Rohden, Sorge,
  Witthaut, and Timme}}]{13powerlong}
\bibinfo{author}{\bibfnamefont{M.}~\bibnamefont{Rohden}},
  \bibinfo{author}{\bibfnamefont{A.}~\bibnamefont{Sorge}},
  \bibinfo{author}{\bibfnamefont{D.}~\bibnamefont{Witthaut}}, \bibnamefont{and}
  \bibinfo{author}{\bibfnamefont{M.}~\bibnamefont{Timme}},
  \bibinfo{journal}{Chaos} \textbf{\bibinfo{volume}{24}},
  \bibinfo{pages}{013123} (\bibinfo{year}{2014}).

\bibitem[{\citenamefont{Nishikawa and Motter}(2015)}]{Nish15}
\bibinfo{author}{\bibfnamefont{T.}~\bibnamefont{Nishikawa}} \bibnamefont{and}
  \bibinfo{author}{\bibfnamefont{A.~E.} \bibnamefont{Motter}},
  \bibinfo{journal}{New Journal of Physics} \textbf{\bibinfo{volume}{17}},
  \bibinfo{pages}{015012} (\bibinfo{year}{2015}).

\bibitem[{\citenamefont{Machowski et~al.}(2008)\citenamefont{Machowski, Bialek,
  and Bumby}}]{Mach08}
\bibinfo{author}{\bibfnamefont{J.}~\bibnamefont{Machowski}},
  \bibinfo{author}{\bibfnamefont{J.}~\bibnamefont{Bialek}}, \bibnamefont{and}
  \bibinfo{author}{\bibfnamefont{J.}~\bibnamefont{Bumby}},
  \emph{\bibinfo{title}{Power system dynamics, stability and control}}
  (\bibinfo{publisher}{John Wiley \& Sons}, \bibinfo{address}{New York},
  \bibinfo{year}{2008}).

\bibitem[{\citenamefont{Filatrella et~al.}(2008)\citenamefont{Filatrella,
  Nielsen, and Pedersen}}]{Fila08}
\bibinfo{author}{\bibfnamefont{G.}~\bibnamefont{Filatrella}},
  \bibinfo{author}{\bibfnamefont{A.~H.} \bibnamefont{Nielsen}},
  \bibnamefont{and} \bibinfo{author}{\bibfnamefont{N.~F.}
  \bibnamefont{Pedersen}}, \bibinfo{journal}{Eur. Phys. J. B}
  \textbf{\bibinfo{volume}{61}}, \bibinfo{pages}{485} (\bibinfo{year}{2008}).

\bibitem[{\citenamefont{Bergen and Hill}(1981)}]{Berg81}
\bibinfo{author}{\bibfnamefont{A.~R.} \bibnamefont{Bergen}} \bibnamefont{and}
  \bibinfo{author}{\bibfnamefont{D.~J.} \bibnamefont{Hill}},
  \bibinfo{journal}{IEEE Transactions on Power Apparatus and Systems}
  \textbf{\bibinfo{volume}{PAS-100}}, \bibinfo{pages}{25}
  (\bibinfo{year}{1981}).

\bibitem[{\citenamefont{Wood et~al.}(2013)\citenamefont{Wood, Wollenberg, and
  Shebl\'e}}]{Wood13}
\bibinfo{author}{\bibfnamefont{A.~J.} \bibnamefont{Wood}},
  \bibinfo{author}{\bibfnamefont{B.~F.} \bibnamefont{Wollenberg}},
  \bibnamefont{and} \bibinfo{author}{\bibfnamefont{G.~B.}
  \bibnamefont{Shebl\'e}}, \emph{\bibinfo{title}{Power Generation, Operation
  and Control}} (\bibinfo{publisher}{John Wiley \& Sons}, \bibinfo{address}{New
  York}, \bibinfo{year}{2013}).

\bibitem[{\citenamefont{Ronellenfitsch
  et~al.}(2016{\natexlab{a}})\citenamefont{Ronellenfitsch, Timme, and
  Witthaut}}]{16ptdf}
\bibinfo{author}{\bibfnamefont{H.}~\bibnamefont{Ronellenfitsch}},
  \bibinfo{author}{\bibfnamefont{M.}~\bibnamefont{Timme}}, \bibnamefont{and}
  \bibinfo{author}{\bibfnamefont{D.}~\bibnamefont{Witthaut}},
  \bibinfo{journal}{IEEE Transactions on Power Systems}
  \textbf{\bibinfo{volume}{PP}} (\bibinfo{year}{2016}{\natexlab{a}}).

\bibitem[{\citenamefont{{D.Manik, D. Witthaut, B. Scha\"fer, M. Matthiae, A.
  Sorge, M. Rohden, E. Katifori and M. Timme}}(2014)}]{14bifurcation}
\bibinfo{author}{\bibnamefont{{D.Manik, D. Witthaut, B. Scha\"fer, M. Matthiae,
  A. Sorge, M. Rohden, E. Katifori and M. Timme}}}, \bibinfo{journal}{Eur.
  Phys. J. Special Topics} \textbf{\bibinfo{volume}{223}},
  \bibinfo{pages}{2527} (\bibinfo{year}{2014}).

\bibitem[{\citenamefont{Skardal et~al.}(2014)\citenamefont{Skardal, Taylor, and
  Sun}}]{Skar14}
\bibinfo{author}{\bibfnamefont{P.~S.} \bibnamefont{Skardal}},
  \bibinfo{author}{\bibfnamefont{D.}~\bibnamefont{Taylor}}, \bibnamefont{and}
  \bibinfo{author}{\bibfnamefont{J.}~\bibnamefont{Sun}},
  \bibinfo{journal}{Phys. Rev. Lett.} \textbf{\bibinfo{volume}{113}},
  \bibinfo{pages}{144101} (\bibinfo{year}{2014}).

\bibitem[{\citenamefont{Fiedler}(1973)}]{Fied73}
\bibinfo{author}{\bibfnamefont{M.}~\bibnamefont{Fiedler}},
  \bibinfo{journal}{Czechoslovak Mathematical Journal}
  \textbf{\bibinfo{volume}{23}}, \bibinfo{pages}{298} (\bibinfo{year}{1973}).

\bibitem[{\citenamefont{Fortunato}(2010)}]{Fort10}
\bibinfo{author}{\bibfnamefont{S.}~\bibnamefont{Fortunato}},
  \bibinfo{journal}{Physics Reports} \textbf{\bibinfo{volume}{486}},
  \bibinfo{pages}{75} (\bibinfo{year}{2010}).

\bibitem[{\citenamefont{Simonsen et~al.}(2008)\citenamefont{Simonsen, Buzna,
  Peters, Bornholdt, and Helbing}}]{Simo08}
\bibinfo{author}{\bibfnamefont{I.}~\bibnamefont{Simonsen}},
  \bibinfo{author}{\bibfnamefont{L.}~\bibnamefont{Buzna}},
  \bibinfo{author}{\bibfnamefont{K.}~\bibnamefont{Peters}},
  \bibinfo{author}{\bibfnamefont{S.}~\bibnamefont{Bornholdt}},
  \bibnamefont{and} \bibinfo{author}{\bibfnamefont{D.}~\bibnamefont{Helbing}},
  \bibinfo{journal}{Phys. Rev. Lett.} \textbf{\bibinfo{volume}{100}},
  \bibinfo{pages}{218701} (\bibinfo{year}{2008}).

\bibitem[{\citenamefont{Ballentine}(1998)}]{Ball98}
\bibinfo{author}{\bibfnamefont{L.~E.} \bibnamefont{Ballentine}},
  \emph{\bibinfo{title}{Quantum Mechanics -- A Modern Development}}
  (\bibinfo{publisher}{World Scientific}, \bibinfo{address}{Singapore},
  \bibinfo{year}{1998}).

\bibitem[{\citenamefont{Newman}(2005)}]{Newm05}
\bibinfo{author}{\bibfnamefont{M.~E.~J.} \bibnamefont{Newman}},
  \bibinfo{journal}{Social Networks} \textbf{\bibinfo{volume}{27}},
  \bibinfo{pages}{39} (\bibinfo{year}{2005}).

\bibitem[{\citenamefont{Brandes and Fleischer}(2005)}]{Brandes2005}
\bibinfo{author}{\bibfnamefont{U.}~\bibnamefont{Brandes}} \bibnamefont{and}
  \bibinfo{author}{\bibfnamefont{D.}~\bibnamefont{Fleischer}},
  \emph{\bibinfo{title}{Centrality Measures Based on Current Flow}}
  (\bibinfo{publisher}{Springer Berlin Heidelberg}, \bibinfo{address}{Berlin,
  Heidelberg}, \bibinfo{year}{2005}), pp. \bibinfo{pages}{533--544}.

\bibitem[{\citenamefont{Palacios}(2001)}]{palacios2001closed}
\bibinfo{author}{\bibfnamefont{J.~L.} \bibnamefont{Palacios}},
  \bibinfo{journal}{International Journal of Quantum Chemistry}
  \textbf{\bibinfo{volume}{81}}, \bibinfo{pages}{135} (\bibinfo{year}{2001}).

\bibitem[{\citenamefont{Hutcheon and Bialek}(2013)}]{Hutc13}
\bibinfo{author}{\bibfnamefont{N.}~\bibnamefont{Hutcheon}} \bibnamefont{and}
  \bibinfo{author}{\bibfnamefont{J.~W.} \bibnamefont{Bialek}},
  \bibinfo{journal}{2013 IEEE Grenoble Conference PowerTech, POWERTECH 2013}
  pp. \bibinfo{pages}{1--5} (\bibinfo{year}{2013}).

\bibitem[{\citenamefont{Jung and Kettemann}(2016)}]{jungpre2016}
\bibinfo{author}{\bibfnamefont{D.}~\bibnamefont{Jung}} \bibnamefont{and}
  \bibinfo{author}{\bibfnamefont{S.}~\bibnamefont{Kettemann}},
  \bibinfo{journal}{Phys. Rev. E} \textbf{\bibinfo{volume}{94}},
  \bibinfo{pages}{012307} (\bibinfo{year}{2016}).

\bibitem[{\citenamefont{Pecora and Carroll}(1998)}]{Peco98}
\bibinfo{author}{\bibfnamefont{L.~M.} \bibnamefont{Pecora}} \bibnamefont{and}
  \bibinfo{author}{\bibfnamefont{T.~L.} \bibnamefont{Carroll}},
  \bibinfo{journal}{Phys. Rev. Lett.} \textbf{\bibinfo{volume}{80}},
  \bibinfo{pages}{2109} (\bibinfo{year}{1998}).

\bibitem[{\citenamefont{Menck et~al.}(2013)\citenamefont{Menck, Heitzig,
  Marwan, and Kurths}}]{Menc13}
\bibinfo{author}{\bibfnamefont{P.~J.} \bibnamefont{Menck}},
  \bibinfo{author}{\bibfnamefont{J.}~\bibnamefont{Heitzig}},
  \bibinfo{author}{\bibfnamefont{N.}~\bibnamefont{Marwan}}, \bibnamefont{and}
  \bibinfo{author}{\bibfnamefont{J.}~\bibnamefont{Kurths}},
  \bibinfo{journal}{Nature Physics} \textbf{\bibinfo{volume}{9}},
  \bibinfo{pages}{89} (\bibinfo{year}{2013}).

\bibitem[{\citenamefont{Menck et~al.}(2014)\citenamefont{Menck, Heitzig,
  Kurths, and Schellnhuber}}]{Menc14}
\bibinfo{author}{\bibfnamefont{P.~J.} \bibnamefont{Menck}},
  \bibinfo{author}{\bibfnamefont{J.}~\bibnamefont{Heitzig}},
  \bibinfo{author}{\bibfnamefont{J.}~\bibnamefont{Kurths}}, \bibnamefont{and}
  \bibinfo{author}{\bibfnamefont{H.~J.} \bibnamefont{Schellnhuber}},
  \bibinfo{journal}{Nature Comm.} \textbf{\bibinfo{volume}{5}},
  \bibinfo{pages}{3969} (\bibinfo{year}{2014}).

\bibitem[{\citenamefont{Albert et~al.}(2000)\citenamefont{Albert, Jeong, and
  Barab{\'a}si}}]{Albe00}
\bibinfo{author}{\bibfnamefont{R.}~\bibnamefont{Albert}},
  \bibinfo{author}{\bibfnamefont{H.}~\bibnamefont{Jeong}}, \bibnamefont{and}
  \bibinfo{author}{\bibfnamefont{A.}~\bibnamefont{Barab{\'a}si}},
  \bibinfo{journal}{Nature} \textbf{\bibinfo{volume}{406}},
  \bibinfo{pages}{378} (\bibinfo{year}{2000}).

\bibitem[{\citenamefont{Cohen et~al.}(2001)\citenamefont{Cohen, Erez, ben
  Avraham, and Havlin}}]{Cohe01}
\bibinfo{author}{\bibfnamefont{R.}~\bibnamefont{Cohen}},
  \bibinfo{author}{\bibfnamefont{K.}~\bibnamefont{Erez}},
  \bibinfo{author}{\bibfnamefont{D.}~\bibnamefont{ben Avraham}},
  \bibnamefont{and} \bibinfo{author}{\bibfnamefont{S.}~\bibnamefont{Havlin}},
  \bibinfo{journal}{Phys. Rev. Lett.} \textbf{\bibinfo{volume}{86}},
  \bibinfo{pages}{3682} (\bibinfo{year}{2001}).

\bibitem[{\citenamefont{Motter and Lai}(2002)}]{Mott02}
\bibinfo{author}{\bibfnamefont{A.~E.} \bibnamefont{Motter}} \bibnamefont{and}
  \bibinfo{author}{\bibfnamefont{Y.-C.} \bibnamefont{Lai}},
  \bibinfo{journal}{Phys. Rev. E} \textbf{\bibinfo{volume}{66}},
  \bibinfo{pages}{065102} (\bibinfo{year}{2002}).

\bibitem[{\citenamefont{Albert et~al.}(2004)\citenamefont{Albert, Albert, and
  Nakarado}}]{Albe04}
\bibinfo{author}{\bibfnamefont{R.}~\bibnamefont{Albert}},
  \bibinfo{author}{\bibfnamefont{I.}~\bibnamefont{Albert}}, \bibnamefont{and}
  \bibinfo{author}{\bibfnamefont{G.~L.} \bibnamefont{Nakarado}},
  \bibinfo{journal}{Phys. Rev. E} \textbf{\bibinfo{volume}{69}},
  \bibinfo{pages}{025103} (\bibinfo{year}{2004}).

\bibitem[{\citenamefont{Buldyrev et~al.}(2010)\citenamefont{Buldyrev, Parshani,
  Paul, Stanley, and Havlin}}]{Buld10}
\bibinfo{author}{\bibfnamefont{S.~V.} \bibnamefont{Buldyrev}},
  \bibinfo{author}{\bibfnamefont{R.}~\bibnamefont{Parshani}},
  \bibinfo{author}{\bibfnamefont{G.}~\bibnamefont{Paul}},
  \bibinfo{author}{\bibfnamefont{H.~E.} \bibnamefont{Stanley}},
  \bibnamefont{and} \bibinfo{author}{\bibfnamefont{S.}~\bibnamefont{Havlin}},
  \bibinfo{journal}{Nature} \textbf{\bibinfo{volume}{464}},
  \bibinfo{pages}{1025} (\bibinfo{year}{2010}).

\bibitem[{\citenamefont{Fairley}(2004)}]{Fair04}
\bibinfo{author}{\bibfnamefont{P.}~\bibnamefont{Fairley}},
  \bibinfo{journal}{IEEE Spectrum} \textbf{\bibinfo{volume}{41}},
  \bibinfo{pages}{22} (\bibinfo{year}{2004}).

\bibitem[{\citenamefont{Pourbeik et~al.}(2006)\citenamefont{Pourbeik, Kundur,
  and Taylor}}]{Pour06}
\bibinfo{author}{\bibfnamefont{P.}~\bibnamefont{Pourbeik}},
  \bibinfo{author}{\bibfnamefont{P.}~\bibnamefont{Kundur}}, \bibnamefont{and}
  \bibinfo{author}{\bibfnamefont{C.}~\bibnamefont{Taylor}},
  \bibinfo{journal}{IEEE Power and Energy Magazine}
  \textbf{\bibinfo{volume}{4}}, \bibinfo{pages}{22} (\bibinfo{year}{2006}).

\bibitem[{\citenamefont{Woodbury}()}]{Wood50}
\bibinfo{author}{\bibfnamefont{M.~A.} \bibnamefont{Woodbury}},
  \emph{\bibinfo{title}{Inverting modified matrices}},
  \bibinfo{howpublished}{Memorandum report 42, Statistical Research Group,
  Princeton University}.

\bibitem[{\citenamefont{Ronellenfitsch
  et~al.}(2016{\natexlab{b}})\citenamefont{Ronellenfitsch, Manik, H{\"o}rsch,
  Brown, and Witthaut}}]{LODF2016}
\bibinfo{author}{\bibfnamefont{H.}~\bibnamefont{Ronellenfitsch}},
  \bibinfo{author}{\bibfnamefont{D.}~\bibnamefont{Manik}},
  \bibinfo{author}{\bibfnamefont{J.}~\bibnamefont{H{\"o}rsch}},
  \bibinfo{author}{\bibfnamefont{T.}~\bibnamefont{Brown}}, \bibnamefont{and}
  \bibinfo{author}{\bibfnamefont{D.}~\bibnamefont{Witthaut}},
  \bibinfo{journal}{arXiv preprint arXiv:1606.07276}
  (\bibinfo{year}{2016}{\natexlab{b}}).

\bibitem[{\citenamefont{Witthaut et~al.}(2016)\citenamefont{Witthaut, Rohden,
  Zhang, Hallerberg, and Timme}}]{16critical}
\bibinfo{author}{\bibfnamefont{D.}~\bibnamefont{Witthaut}},
  \bibinfo{author}{\bibfnamefont{M.}~\bibnamefont{Rohden}},
  \bibinfo{author}{\bibfnamefont{X.}~\bibnamefont{Zhang}},
  \bibinfo{author}{\bibfnamefont{S.}~\bibnamefont{Hallerberg}},
  \bibnamefont{and} \bibinfo{author}{\bibfnamefont{M.}~\bibnamefont{Timme}},
  \bibinfo{journal}{Phys.~Rev.~Lett.} \textbf{\bibinfo{volume}{116}},
  \bibinfo{pages}{138701} (\bibinfo{year}{2016}).

\bibitem[{\citenamefont{Egan}(1975)}]{Egan75}
\bibinfo{author}{\bibfnamefont{J.~P.} \bibnamefont{Egan}},
  \emph{\bibinfo{title}{Signal Detection Theory and ROC Analysis}}
  (\bibinfo{publisher}{Academic Press}, \bibinfo{year}{1975}).

\end{thebibliography}
\bibliographystyle{apsrev}

\end{document}